\documentclass[11pt]{article}
\usepackage{mathtools,amsthm,amsmath,amssymb,authblk}
\usepackage{geometry,graphicx,float}
\usepackage{xcolor}
\usepackage{mdframed}
\usepackage{stmaryrd}
\usepackage{booktabs}   
\usepackage{hyperref}
\DeclareMathOperator{\argmin}{argmin}

\newcommand{\asinh}{\operatorname{asinh}}

\usepackage{soul}
\title{Machine-learned acceleration of earthquake sequence simulations}
\author{Reo Yanagi, Taeho Kim\footnote{Now at Stanford University}, Kaushik Bhattacharya, and Nadia Lapusta}
\affil{California Institute of Technology}

\date{Nov. 16, 2026}
\begin{document}

\maketitle

\section*{Abstract}
Earthquake sequence simulations involve slow interseismic processes and rapid dynamic ruptures. During interseismic periods, a numerical solver can take large time steps efficiently. During seismic periods, in contrast, the solver must drastically reduce the time step, and elastodynamic wave-mediated memory effects become significant; long-term simulations are therefore computationally expensive. We investigate the use of neural operators to accelerate earthquake sequence simulations. We train a Fourier neural operator (FNO) on data from high-fidelity, fully dynamic simulations to learn a jump map that advances the system state from before a seismic rupture to after it. We propose a hybrid modeling framework in which a conventional quasi-dynamic solver simulates the slow interseismic evolution, and the trained FNO replaces the solver during the computationally intensive seismic periods with wave-mediated inertial effects. The hybrid model reproduces key long-term statistics of fully dynamic simulations, including distributions of event magnitudes and recurrence times. Furthermore, it provides a much better approximation to the fully dynamic simulations than the commonly used quasi-dynamic approximation at a reduced computational expense.

\section{Introduction}
Earthquakes typically occur as dynamic shear ruptures on pre-existing faults under slow loading and involve long interseismic periods of slow, quasi-static deformation punctuated by rapid fault slip~\cite{avouac2015geodetic,harris2017large}. The slow and fast slip processes co-exist and interact; in part, slow slip modifies the fault stress and other conditions before dynamic rupture and can act as earthquake nucleation~\cite{avouac2015geodetic,harris2017large}. To study these interactions and earthquake source physics, simulations of sequences of earthquakes and aseismic slip (SEAS) have been developed~\cite{rice_spatiotemporal_1993,benzion1997dynamic,lapusta2000,erickson2020community,jiang2022community}. Such simulations are challenging since they need to resolve multiscale processes in space and time, from the meter-scale, sub-second evolution of stress and slip at the rupture front to fault dimensions in kilometers and loading periods of hundreds of years. 

One of the main challenges in SEAS simulations is the inclusion of inertial, wave-mediated stress changes during rapid earthquake rupture~\cite{benzion1997dynamic, lapusta2000,thomas_quasidynamic_2014}. Once earthquake rupture begins, the inertial, wave-mediated stress changes carried by seismic waves are typically much larger than the static stress changes, and they significantly affect rupture propagation. That is why such wave-mediated stress changes are routinely included in simulations of single dynamic events, often called dynamic rupture simulations~\cite{olsen1997three, harris2009scec, taufiqurrahman2023dynamics}.  Tracking wave-mediated stress transfers can become prohibitively expensive in SEAS simulations that involve multiple earthquake events, because it requires both significant additional computation and massive storage, and that is why inertial effects are commonly neglected and replaced by quasi-dynamic approximations \cite{rice_spatiotemporal_1993} and other simplifications~\cite{richardsdinger2012rsqsim}. However, quasi-dynamic approximations can significantly modify earthquake sequence simulations\cite{thomas_quasidynamic_2014,agajanian2026supershear}. This motivates us to accelerate fully dynamic (FD) simulations of earthquake sequences.

At the same time, machine learning methods for scientific computing have developed significantly. Neural networks are expressive functions, and they can approximate any continuous function on a compact subset of Euclidean space to any accuracy \cite{cybenko1989approximation}. However, the fault dynamics that we seek to simulate are described by partial differential equations (PDEs). The solution operators of PDEs are maps between infinite-dimensional function spaces. Neural operators generalize neural networks to such maps by replacing weight matrices with integral operators that have learned kernels; they have universal approximation theorems between function spaces \cite{kovachki2023neural_operator}. There are several operator learning methods, including DeepONet \cite{lu2021deeponet}, PCA-Net \cite{bhattacharya2021model}, Fourier neural operator (FNO) \cite{li2021fourier}, and random features \cite{nelsen2021random}.
For reviews, see \cite{kovachki2024operator, boulle2024mathematical, bhattacharya2026multiscale}.  Machine learning methods have been applied to dynamic rupture simulations~\cite{gong2025rupturenet2d, tainpakdipat2025fourier}. These studies aim to train a neural network to reproduce the outcomes of a single dynamic rupture.

In this study, we use neural operators to accelerate earthquake sequence simulations. To that end, we develop a hybrid approach in which conventional simulations are used to simulate slow interseismic periods while neural operators replace the computationally expensive dynamic rupture periods. Specifically, we use an FNO to learn the jump map from the fault states before earthquake rupture to the fault states after the rupture. Such a hybrid approach is advantageous in two ways. First, since the inertial, wave-mediated effects are negligible before and after the earthquake rupture, the jump map does not depend on history, i.e., it is Markovian, which makes this problem suitable for neural operators which are essentially Markovian maps.  Second, the simulation of the interseismic periods is quasi-static and hence relatively inexpensive. 
Mathematically, our method can be viewed as learning the Poincaré map on the Poincaré section defined by the velocity threshold that separates the fast and slow phases. We illustrate this view with the spring-slider system described by a system of ordinary differential equations (ODEs) (Appendix~\ref{sec:sprint-slider}). 
We find that the FNO surrogate can indeed learn the earthquake jump map.  The resulting hybrid simulations recover several types of long-term statistics from the fully dynamic earthquake sequence simulations, including the distributions of earthquake magnitudes and recurrence times. They also approximate the higher-order statistics, such as the recurrence times of larger events and the transition probabilities between smaller and larger events.

The paper is organized as follows. Section~\ref{sec:eq-modeling} describes the fault model, its FD and QD simulations, the definition of an earthquake, and the invariant statistics. Section~\ref{sec:operator-learning} reviews operator learning and the FNO. Section~\ref{sec:hybrid-method} defines the earthquake jump map, develops its FNO surrogate, and proposes the algorithm for the hybrid simulator. Section~\ref{sec:results} presents the comparison between the hybrid, fully dynamic, and quasi-dynamic simulations.  We conclude and discuss future work in Section~\ref{sec:conclusion}. Appendix~\ref{sec:technical} gives technical details, and Appendix~\ref{sec:sprint-slider} applies the method to the spring-slider system.

\newpage
\section{Background on Earthquake Modeling}\label{sec:eq-modeling}

\subsection{Earthquake Model}
\label{sec:model}
\begin{figure}
    \centering
    \includegraphics[width=1\linewidth]{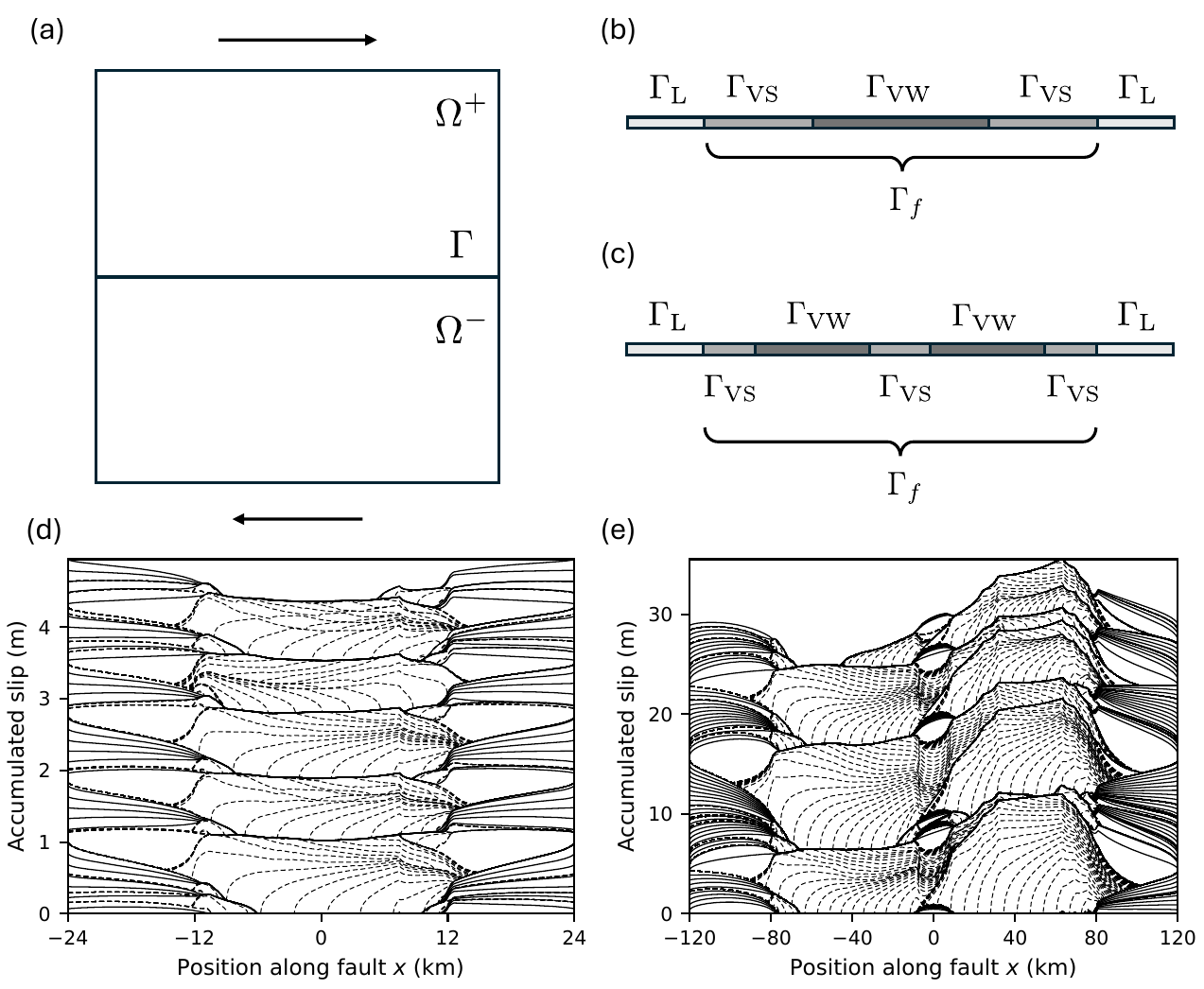}
    \caption{Model geometry and representative simulations. (a) Two-dimensional elastodynamic space $\Omega = \mathbb{T}^1_L \times \mathbb{R}$ split by the planar interface (fault) $\Gamma$ into $\Omega^{+}$ and $\Omega^{-}$. (b) Partition of $\Gamma$ into velocity-weakening (VW), velocity-strengthening (VS), and loading regions $\Gamma_{\mathrm{L}}$ for the homogeneous case. In the loading regions $\Gamma_{\mathrm{L}}$, tectonic plate velocity $V_{\mathrm{pl}}$ is applied across the interface. (c) Corresponding partition of $\Gamma$ for the heterogeneous case. (d,e) Slip accumulation over the portion of the fault close to the VW regions for the homogeneous and heterogeneous cases, respectively. Solid lines show slip accumulation every 5 years in (d) and 10 years in (e), and dashed lines show slip accumulation every 1 s in (d) and 2 s in (e) during earthquake events.}
    \label{fig1}
\end{figure}

We consider the model problem commonly used in earthquake sequence simulations \cite{lapusta2000}, with a one-dimensional frictional interface governed by a rate-and-state friction law in an unbounded, homogeneous, isotropic, linear-elastic solid under Mode III (antiplane) deformation (Figure \ref{fig1}).

\paragraph{Domain} We consider a two-dimensional body $\Omega := \mathbb{T}^1_L\times\mathbb{R}$, which is unbounded in the vertical direction $y$ and periodic in $x$ with period $L$. $\Omega$ is separated into $\Omega^+$ and $\Omega^-$ by the horizontal interface, $\Gamma := \{(x,0): x\in[-L/2,L/2]\},
$ which represents the fault (Figure \ref{fig1} (a)). We separate $\Gamma$ into a frictional region ($\Gamma_f$) and a loading region ($\Gamma_L$) (the details of this segmentation are discussed below).

\paragraph{Governing equations} \label{sec:ge}
For anti-plane shear, the displacement $u: \Omega \times [0,\infty) \to {\mathbb R}$ is a scalar.  It satisfies the equations of elastodynamics which reduce to the scalar wave equation in each half space with density $\rho>0$ and shear modulus $\mu>0$.  We use the following common definitions: the slip on the interface, $\delta(x,t)$, is the jump in the displacement across the interface, $\delta(x,t) := \llbracket u \rrbracket(x,t) = u^{+}(x,0,t) - u^{-}(x,0,t)$,  where $u^{\pm}(x,y,t)$ are the displacement fields in $\Omega^{\pm}$; the slip rate is $V(x,t) := \partial_t \delta(x,t)$; the scalar shear traction is $\tau(x,t)$ on $\Gamma$; it follows from elasticity that $\tau(x,t) = \mu\,\partial_y u^\pm(x,0^\pm,t)$.

One can obtain the following integro-differential equation that relates the traction across the interface to the history of the slip \cite{perrin1995}:
\begin{equation}\label{eq:bie}
  \tau(x,t) = \tau_0(x) + \phi[\delta,V](x,t)
  - \frac{\mu}{2c_s} V(x,t),
\end{equation}
where $\tau_0$ is the initial shear stress distribution and $c_s = \sqrt{\mu/\rho}$ is the shear-wave speed. The functional $\phi[\delta,V]$ represents the elastodynamic stress transfer due to slip and slip rate, which we can express in Fourier space,
\begin{equation}\label{eq:phi}
  \widehat{\phi}(k,t)
  = -\frac{\mu |k|}{2}\,\widehat{\delta}(k,t)
  + \frac{\mu |k|}{2} \int_0^t
      W(|k|c_s t')\,\widehat{V}(k,t-t')\,\mathrm{d}t',
\end{equation}
where (\,$\widehat{\cdot}\,$) represents the Fourier transformed variable. The first term describes the instantaneous static response, while 
the convolution term captures the wave-memory effect arising from past slip-rate history, where $W$ is the anti-plane stress kernel from elastodynamics, $W(p) = \int_p^\infty \frac{J_1(\xi)}{\xi}\,\mathrm{d}\xi$, and $J_1$ is the Bessel function of the first kind of order one. This formulation is advantageous because we can now reduce the problem to consider only $\Gamma$.

A constitutive relation that describes the frictional behavior of the interface closes the system of equations.  We use the rate-and-state friction \cite{dieterich1979,ruina1983} which postulates that the frictional resistance depends on the slip rate $V$ and the state variable $\Theta$. The state variable represents the history of slip and evolves as a function of the current values of $V$ and $\Theta$. Hence we write: 
\begin{equation}\label{eq:rs}
\begin{aligned}
    \tau & = F(V, \Theta), \quad
    \partial_t \Theta &= G(V,\Theta)
\end{aligned}.
\end{equation}

By equating \eqref{eq:bie} and \eqref{eq:rs}, and taking the time derivative, the dynamics of the system can be described in terms of the slip velocity $V(x,t)$ and state variable $\Theta(x,t)$ that are governed by the system of equations
\begin{equation}\label{eq:FD}\tag{FD}
\left\{
\begin{aligned}
   \Bigl(\frac{\mu}{2c_s} + \partial_VF(V,\Theta)\Bigr)\partial_t V &= 
\mathcal{K}[V]
- \partial_{\Theta}F(V,\Theta)G(V,\Theta)\qquad 
    && \text{in } \Gamma_f\times(0,T],\\
 \partial_t \Theta &= G(V,\Theta)
    && \text{in } \Gamma_f\times(0,T],\\
V &\equiv V_{\mathrm{pl}}
    && \text{in } \Gamma_{L}\times(0,T],
\end{aligned}
\right.
\end{equation}
where it is assumed that the fault is loaded by continuous motion with $V \equiv V_{\mathrm{pl}}$ over the portions $\Gamma_{L}$ of the interface due to the far-field tectonic plate rate.  The history-dependent operator $\mathcal{K}$ is
defined in Fourier space by
\[
\widehat{\mathcal{K}[V]}(k,t)
=
-\frac{\mu |k|}{2}
\int_0^t
M(k,t-t')\,\widehat{V}(k,t')\,\mathrm{d}t',
\]
where $M(k, t) = \frac{J_1(|k|c_st)}{t}$. We solve \eqref{eq:FD} with initial data  $V(\cdot, 0) = V_0$, $\Theta(\cdot,0)=\Theta_0$.

We refer to this system as the {\it Fully Dynamic (FD)} formulation. A common approximation is called the {\it Quasi-Dynamic (QD)} formulation, in which the history-dependent part of stress transfer in \eqref{eq:phi} is ignored \cite{rice_spatiotemporal_1993,thomas_quasidynamic_2014}. With this approximation, the dynamical system is local in time, so we do not need the history of $V$ to evaluate the system. Specifically, the operator in \eqref{eq:FD} becomes $\mathcal{K}[V] \approx -\frac{\mu}{2}(-\Delta)^{1/2}V$.



\paragraph{Rate-and-state friction law}We use a laboratory-derived form~\cite{dieterich1979, marone1998,ruina1983} for $F$, regularized at zero slip rate~\cite{lapusta2000}, and the Dieterich--Ruina aging law for $G$~\cite{dieterich1979, ruina1983}.
\begin{align}
&F(V,\Theta) = a\sigma_n\,
  \asinh\left(
    \frac{V}{2V^*}
    \exp\left(\frac{f^*+b\Theta}{a}\right)
  \right),
  \label{eq:friction-law}\\
&G(V,\Theta) = \frac{1}{D_{\rm RS}}(V^*e^{-\Theta}-V),
  \label{eq:aging}
\end{align}
where, $\sigma_n$ is the normal stress, $f^*$ is the friction coefficient at the reference slip rate $V^*$, and $a$, $b$, and $D_{\rm RS}$ are rate-and-state parameters, which are functions of position $x$ along the fault. When $a-b<0$, the steady-state friction response is velocity-weakening (VW), and when $a-b>0$, the steady-state friction response is velocity-strengthening (VS).  We typically use a uniform $a$ and a piecewise-constant distribution of $b$. Stability analyses and past simulations show that VW regions are potentially seismogenic, i.e., can accelerate to fast, inertially controlled slip for large enough regions even under slow tectonic-like loading, whereas VS regions respond to slow loading with stable, quasi-static slip~\cite{rice1983stability, dieterich1992nucleation, rice2001rate, rubin2005nucleation}.  The rate-and-state friction formulation has been widely used to simulate sequences of earthquakes and aseismic slip~\cite{rice_spatiotemporal_1993,benzion1997dynamic,lapusta2000,allison2018earthquake,erickson2020community,jiang2022community}, as it can represent both stably slipping and earthquake-producing faults and includes interseismic healing through the state variable.

We study two cases as shown in Figure \ref{fig1}(b,c).  Both have a loading region $\Gamma_\mathrm{L}$ at the edges where we impose the far-field slip velocity $V_{\rm{pl}}$.  In the first case shown in Figure \ref{fig1}(b), the frictional region $\Gamma_{f}$ has three segments with a velocity-weakening (VW) region $\Gamma_\mathrm{VW}$ surrounded by velocity-strengthening (VS) regions $\Gamma_{\mathrm{VS}}$. We call this case ``{\it homogeneous fault}''  since the seismogenic, VW portion of the fault is homogeneous.  In the {\it heterogeneous fault} case shown in Figure \ref{fig1}(c), the VW portion of the fault is separated into two segments by an additional smaller VS region. 

The specific earthquake model parameters used in this study are given in Table \ref{tab:faultparams}.

\begin{table}
    \caption{Earthquake model parameters for the homogeneous and heterogeneous faults.}
    \label{tab:faultparams}
    \centering
    \begin{tabular}{l c c c}
         Parameter & Symbol & Homogeneous \cite{lambert2021scale}& Heterogeneous \cite{kaneko2010towards}\\
         \midrule
         \midrule
         Bulk parameters & & & \\
         Loading rate & $V_\mathrm{pl}$ & $10^{-9}$ m/s & $1.585\times10^{-9}$ m/s \\
         Shear wave speed & $c_s$ & 3,300 m/s & 3,300 m/s \\
         Shear modulus & $\mu$ & 36 GPa & 30 GPa \\
         Effective normal stress & $\sigma_n$ & 25 MPa & 50 MPa \\
         \midrule
         Rate-and-state parameters & & & \\
         Reference slip velocity & $V_*$ & $10^{-6}$ m/s & $10^{-6}$ m/s \\
         Reference friction coefficient & $f_*$ & 0.6 & 0.6 \\
         Rate-and-state direct effect (uniform) & $a$ & 0.010 & 0.010 \\
         Rate-and-state evolution effect (VW) & $b$ & 0.015 & 0.015 \\
         Rate-and-state evolution effect (VS) & $b$ & 0.003 & 0.005 \\
         Characteristic length & $D_{\rm{RS}}$ & 0.001 m & 0.008 m \\
         \midrule
         Length scales of fault & & & \\
         Fault length & $\ell$ & 72 km & 480 km \\
         Frictional domain & $\ell_{\mathrm{fr}}$ & 48 km & 240 km \\
         Velocity-weakening region & $\ell_{\mathrm{VW}}$ & 24 km & $2\times72.5$ km \\
         Central velocity-strengthening patch & $\ell_{\mathrm{VS}}$ & -- & 15 km \\
         \midrule
         \midrule
    \end{tabular}
\end{table}

\subsection{Earthquake  Simulations}\label{sec:earthquake-simulation}
\begin{figure}[t]
    \centering
    \includegraphics[width=0.9\linewidth]{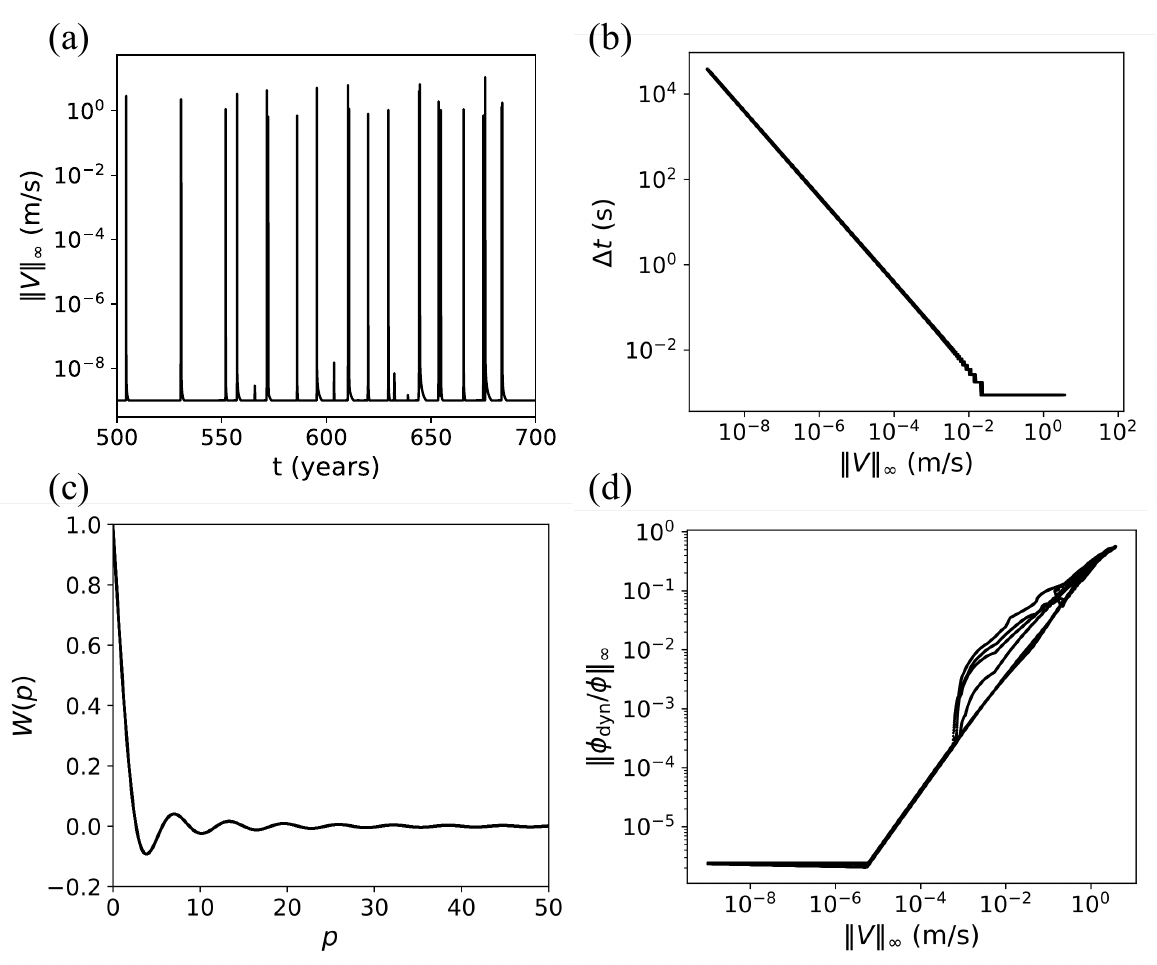}
    \caption{Features of a long-term fully dynamic (FD) simulation in the homogeneous fault case. (a) The time history of the maximum slip velocity over the fault illustrating interseismic periods with interspersed rapid earthquake events. (b) Changes in the adaptive time stepping with the maximum slip velocity over the fault. (c) Decay of the convolution kernel $W$ with its argument $p=|k|c_s t$. (d) The fraction of the stress transfer functional $\phi$ represented by the history-dependent term $\phi_{\mathrm{dyn}}$ as a function of maximum slip velocity over the fault.}
    \label{fig:motivation}
\end{figure}

We use the numerical algorithm developed by Lapusta {\it et al.} \cite{lapusta2000} to simulate the dynamical system \eqref{eq:FD}.  The results of the simulations in terms of slip accumulation on the fault  for a typical time period are shown in Figure~\ref{fig1} (d,e).  We see that the VS regions are accumulating slip steadily, while the VW regions rupture in occasional fast, dynamic events.  The time history of the maximum slip velocity over the fault for the homogeneous case is shown in Figure~\ref{fig:motivation}(a).  There are long periods during which the slip velocity is extremely small, interspersed with short bursts during which the slip velocity can change by many decades (a factor of $10^9$ or more).  The intense bursts are known as {\it seismic events}  or earthquakes, and the intervening slow periods as {\it interseismic} periods.

The complexity and extreme changes in slip velocity mean that the solution of the system \eqref{eq:FD} requires care.  Therefore, the method of Lapusta {\it et al.} \cite{lapusta2000} adopts adaptive time stepping based on the frictional instability and the CFL condition (Figure~\ref{fig:motivation}(b)). During the slow (interseismic) phase, relatively large time steps can be taken, whereas during the rapid (seismic) phase, the time step must be reduced significantly. 

The method also implements a truncation of the dynamic response in time.  The simulated time involves long interseismic periods and therefore directly evaluating the convolution integrals in \eqref{eq:FD} over the entire simulation history would be computationally impractical. However, the wave-mediated effects are negligible over the interseismic period, and hence the convolutions can be truncated. In particular, for a given wavenumber $k$, the convolution kernel $W(|k|c_s t)$ in \eqref{eq:phi} decays like $t^{-3/2}$ (Figure~\ref{fig:motivation}(c)); similarly, the kernel $M(k,t)$ in \eqref{eq:FD} also decays like $t^{-3/2}$. Following \cite{lapusta2000}, we keep only the slip rate history within an elastodynamic time window of length $T_w$, the time for a shear wave to propagate over the velocity-weakening part of the fault, which is a good approximation for the longest possible duration of seismic events.
  
The simulation proceeds as follows.

\begin{mdframed}
{\bf Algorithm P} (fully dynamic simulation~\cite{lapusta2000};details in Appendix ~\ref{sec:technical})

Given $\delta$, $\Theta$, and $V$ at time $t$ and the slip-rate history on $[t-T_w,\,t]$, repeat until $t=T$ where $T$ is the desired simulated time:
\begin{enumerate}
\item {\bf Time step:} Set $\Delta t=\max\bigl\{\Delta t_{\min},\ \min_{x\in\Gamma_f}\zeta(x)\,D_{\rm {RS}}/V(x,t)\bigr\}$, where the factor $\zeta$ comes from the stability analysis of~\cite{lapusta2000} and $\Delta t_{\min}=\Delta x/(\beta c_s)$ is a fraction of the time a shear wave takes to cross a grid cell of width $\Delta x$.
\item {\bf FD step:} Advance $\delta$, $\Theta$, and $V$ to $t+\Delta t$ by one predictor--corrector step of \eqref{eq:FD}.  If $\Delta t<T_w$, the convolution uses the stored slip-rate history; if $\Delta t\geq T_w$, it treats $V$ as constant over the whole window and needs no history.
\end{enumerate}
\end{mdframed}
Algorithm P solves \eqref{eq:FD} at every step and never switches to the QD system; the step size alone separates the slow and fast phases. Although Algorithm P keeps the history-dependent term at every step, this term is small when the slip rate is low.  Let us define
\[
\widehat{\phi}_{\mathrm{dyn}}(k,t)
:= \frac{\mu |k|}{2} \int_0^t
W\!\left(|k|c_s t'\right)\,\widehat{V}\!\left(k,t-t'\right)\,\mathrm{d}t' .
\]
Figure~\ref{fig:motivation}(d) shows the ratio $\|\phi_{\mathrm{dyn}}/\phi\|_\infty$ as a function of
$\|V\|_\infty$: $\phi_{\mathrm{dyn}}$ is negligible when $\|V\|_\infty<10^{-3}\,\mathrm{m/s}$. Note that the QD-FD and FD-QD transition criteria are asymmetric.  We use a lower threshold to restart QD to ensure that all history-dependent parts of the kernel are negligible before we restart QD. (Detailed numerical values of $\|\phi_{\mathrm{dyn}}/\phi\|$ can be found in ~\ref{QDthreshold}.)

\subsection{Earthquake Events}

We define the occurrence of an earthquake event using thresholds on the maximum slip rate.  $\|V(x,t)\|_{\infty}:= \|V(x,t)\|_{L_\infty (\Gamma _f)}$. We choose two thresholds $0<V_2 \leq V_1$. Let us set $t_e^{0}:=0$. For $n \geq 1$, we define the $n$-th event start time $t^n_s$ as the first time after the end of the $(n-1)$-th event, $t_e^{n-1}$, when slip velocity anywhere on the fault exceeds $V_1$:
\begin{equation}\label{eq:event-start}
    t^n_s := \inf \bigl \{ t \in (t_e^{n-1}, T] \; : \; \| V(x,t)\|_\infty \geq V_1\bigr\},
\end{equation}
and, for $t_s^n < \infty$, we define the corresponding event end time as the first subsequent time when the slip velocity everywhere on the fault becomes lower than $V_2$:
\begin{equation}\label{eq:event-end}
    t^n_e := \inf \bigl \{ t \in (t_s^{n}, T] \; : \; \| V(x,t)\|_\infty \leq V_2\bigr\}.
\end{equation}
The $n$-th earthquake event is identified with the time interval $I_n := [t_s^n, t_e^n]$. We choose $V_1$ and $V_2$ to be $10^{-3}$ and $10^{-4}$ m/s, respectively, so that the QD approximation holds outside the events (Figure~\ref{fig:motivation}(d)). 


We define two measures related to an earthquake event.
\paragraph{Event Magnitude}
We define the magnitude $M_n$ of an event $n$ similarly to the definition of moment magnitude for natural events \cite{kanamori2004physics}, but integrating slip over the fault length rather than the fault area:
\begin{equation}\label{eq:Magnitude}
    M_n := \frac{2}{3}\Bigl( \log_{10}\mu \int_{\Gamma} \Delta \delta_n(x)\,dx -9.1\Bigr)
\end{equation}
where 
\begin{equation}
    \Delta \delta_n(x) := \int_{t^n_s}^{t^n_e} V(x,t) \, dt
\end{equation}
is 
the coseismic slip (slip accumulated during the event) at each location $x \in \Gamma$.    

\paragraph{Recurrence time} The recurrence time between events is defined as the start-to-start interval between events, 
\begin{equation}\label{eq:ReccurenceTime}
    T_n := t^{n}_s - t^{n-1}_s, 
\end{equation}
where $\{t_s^n\}_{n \geq 1}$ is the set of the event start times.


\subsection{Invariant Statistics}
\label{sec:invariant-statistics}

\begin{figure}[t]
    \centering
    \includegraphics[width=1\linewidth]{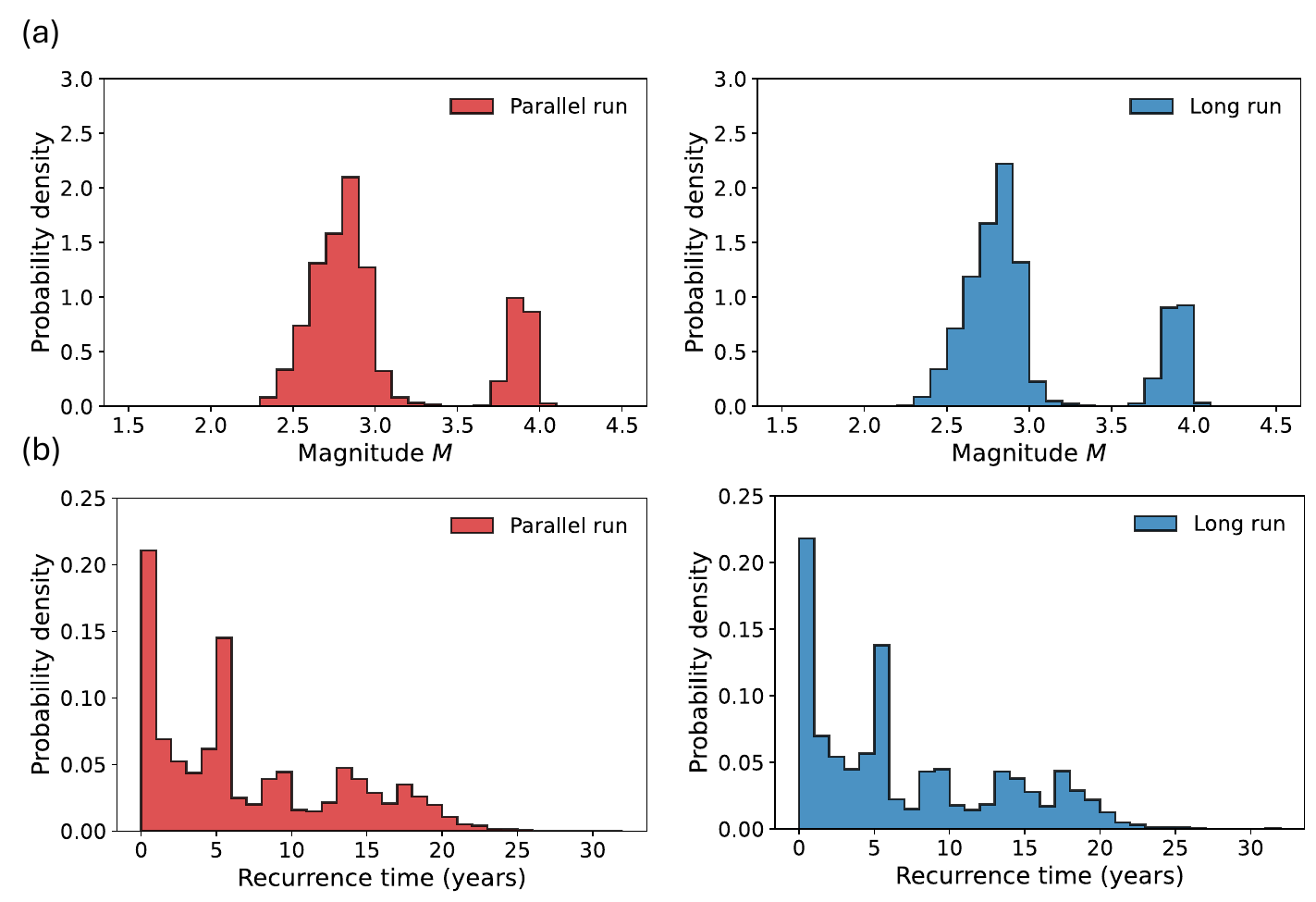}
    \caption{Comparison between (left) a long single simulation (with 6,000 events) and (right) 150 short parallel simulations (with 40 events each) with different randomized initial conditions (see Appendix~\ref{appendix: random-initial-condition}). (a) The probability density of the magnitude distribution. (b) The probability density of the recurrence time.  The two types of simulations result in the same statistics, demonstrating their invariance.}
    \label{fig:invariantmeasure}
\end{figure}

A core postulate of the work is that the dynamical system \eqref{eq:FD} is chaotic with an ergodic global attractor that can support invariant statistics.   To examine this, we perform numerical experiments to obtain 6,000 events by (i) a single long simulation and (ii) an ensemble of shorter simulations run in parallel from randomly chosen initial conditions (in the homogeneous case). For each numerical experiment, we compute the probability densities of the event magnitude and the recurrence time (Figure \ref{fig:invariantmeasure})\footnote{To avoid dependence on the initial conditions, we ignore the first 10 events when computing these statistics. We plot the distribution without skipping the first 10 events and with skipping the first 5 events in Appendix~\ref{sec:initialcondition-dependence}.}. 

In addition to the visual comparison, we use two quantitative measures to quantify the difference between the distributions \cite{bach2026machine}.  The first is the widely used Kullback--Leibler divergence that quantifies the information lost when we approximate a true distribution $P$ with another distribution $Q$, 
\begin{equation}
D_{\mathrm{KL}}(P\, \|\ Q) = \int p(x)\log \left (\frac{p(x)}{q(x)} \right)dx.
\end{equation}
Here, we use the distribution generated by the long sequence as the true distribution.
The second is the $2$-Wasserstein distance that quantifies the optimal transport cost of transforming one distribution to another. In the one-dimensional setting, it is given by the following expression:
\begin{equation}
W_2(p_1, p_2)^2 = \int_0^1 |F_1^{-1}(r)-F_2^{-1}(r)|^2dr,
\end{equation}
where $F_1$ and $F_2$ are the cumulative distribution functions of $p_1$ and $p_2$.

We find that the KL divergence between the two magnitude distributions is 0.0060, and the corresponding divergence for the
recurrence-time distributions is 0.0038.  Similarly, the 2-Wasserstein distance between the
two magnitude distributions is 0.0188, and the corresponding distance for the
recurrence-time distributions is 0.1998.

\section{Background on Operator Learning and Fourier Neural Operator (FNO)}\label{sec:operator-learning}
We seek to bypass the FD simulation of earthquakes by replacing it with a surrogate that advances the state before the earthquake event to the state after the earthquake event.  We do so using operator learning briefly introduced in the following.  We refer the reader to \cite{kovachki2023neural_operator} for details.

Consider an operator (map, process) $\mathcal{G}$ that maps a vector-valued function $u: D \to {\mathbb R}^{d_u}$ defined on a domain $D$ to another vector-valued function $v: D \to {\mathbb R}^{d_v}$.  Note that the dimension $d_u$ of $u$ can be different from that of $v$.  Specifically, 
\begin{equation}
    \mathcal{G}: u(\cdot) \in \mathcal{U}(D;\mathbb{R}^{d_u}) \,\mapsto \, v(\cdot) \in \mathcal{V}(D, \mathbb{R}^{d_v}).
\end{equation}
where $\mathcal U,\mathcal V$ are subsets of some Sobolev spaces.  Consider a situation where we have some data in the form of $n$ input-output pairs of $\mathcal G$, 
\begin{equation}
    \{ u_i, v_i \}_{i=1}^n, \quad v_i = {\mathcal G}(u_i), i = 1, \dots, n.
\end{equation}
We seek to use this data to find an approximation to $\mathcal G$.  We start with a parametrized operator ${\mathcal G}^\text{approx}: \mathcal{U} \times {\mathbb R}^p \to \mathcal{V} $ that depends on $p$ parameters, and find the best choice of parameters to fit the data:
\begin{equation}
    p^* = \argmin_p \sum_i || {\mathcal G}^\text{approx} (u_i; p) - v_i ||_{\mathcal V}.
\end{equation}

The natural choice is a parametrized operator that is highly expressive so that we can approximate a large class of operators.  In this work, we use a  Fourier neural operator (FNO)~\cite{li2021fourier, kovachki2021universal_fno} that has been widely used in a variety of problems.  There is also a universal approximation theorem for continuous operators between Sobolev spaces on the torus \cite{kovachki2021universal_fno}. An FNO is modeled after a neural network, but keeps both spatial and spectral information using Fourier transforms.   An $L$-layer FNO is then defined as a composition:
\begin{equation}
    \mathcal{N}(u) := Q \circ \mathcal{L}_L \circ \mathcal{L}_{L-1}\circ \cdots \circ \mathcal{L}_1 \circ R(u),
\end{equation}
where $R$ is a lifting operator that maps the input channels to a higher-dimensional latent width $d_l$, $Q$ is a projection operator back to $d_v$ channels, and each Fourier layer has the form
\begin{equation}
    (\mathcal{L_\ell} w)(x) = \sigma \left(W_\ell w(x) +b_\ell(x) + \mathcal{F}^{-1} \big( P_\ell (k) \mathcal{F}(w)(k)\big)(x)  \right)
\end{equation}
where $\sigma$ is a nonlinear map (activation), $W_\ell$ is a linear map, and the term  $\mathcal{F}^{-1} \big( P_\ell (k) \mathcal{F}(w)(k)\big)$ defines a nonlocal mapping via multiplication in Fourier space by the learned matrix-valued parameter $P_\ell (k)$. 

\section{Operator Learning and Proposed Hybrid Algorithm}\label{sec:hybrid-method}

As discussed in Section \ref{sec:earthquake-simulation}, the simulation of earthquake events is expensive due to both the history dependence of stress changes in this period and the need to use small time steps.  Therefore, we seek to replace the portions of the simulation that compute dynamic events with a surrogate that is trained from data gathered using multiple short simulations.  We use a Fourier neural operator (FNO) as a surrogate.

\subsection{The earthquake jump map}
Let $U(x,t) = (V(x,t), \Theta(x,t))$ represent our state space. We define the earthquake jump as the map that advances the state of the fault from the one at the start of an earthquake event to the one at the end of it:
\begin{equation}
      \mathcal{S}:U(\cdot,t_s^n) \mapsto U(\cdot,t_e^n).
\end{equation} 


\subsection{The FNO surrogate for the earthquake jump map}\label{section:FNO}

We seek an FNO surrogate $\mathcal{S}^\text{FNO}$ to approximate the earthquake jump map $\mathcal{S}$. We describe the particular architecture and training procedure in this section.

\paragraph{Dataset} In order to train $\mathcal{S}^\text{FNO}$, we require paired samples $\bigl(U(\cdot,t_s^n),\,U(\cdot,t_e^n)\bigr)$ of pre- and post-event states  from the conventional simulator.  For both homogeneous and heterogeneous fault cases, we generate 20,000 paired samples of pre- and post-event states and split them into 14,000 for training, 2,000 for validation during training, and 4,000 for testing. 

As discussed in Section~\ref{sec:invariant-statistics}, an ensemble of shorter simulations provides representative samples of the long-term simulation. Therefore, we generate training data from an ensemble of short, independent simulations initialized from random initial conditions (see Appendix~\ref{appendix: random-initial-condition}). Each run yields multiple events.  We prefer this approach to one long simulation since the dataset can be collected efficiently in parallel. 

\paragraph{FNO model}
We use a 1D FNO with $\tilde{a} \in \mathbb{R}^{N \times d_a}$ as the normalized input, $\tilde u \in \mathbb{R}^{N \times d_u}$ as the normalized output, and 4 Fourier layers:

\begin{equation}
\begin{aligned}
v_0 &= \sigma(P\tilde{a}), && P : \mathbb{R}^{d_a} \to \mathbb{R}^{d_v}\\
v_{l+1} &= \sigma(W_l v_l + \mathcal{K}_l v_l), && l = 0,\dots, 3, \\
\tilde{u} &= Q_2\,\sigma(Q_1 v_4),
&& Q_1: \mathbb{R}^{d_v} \to \mathbb{R}^{128},
\quad Q_2 : \mathbb{R}^{128} \to \mathbb{R}^{d_u},
\end{aligned}
\end{equation}
where the lifting map $P$, the projection maps $Q_1$ and $Q_2$, and the channel mixing $W_l$ are pointwise affine; the latent width $d_v=64$; and $\sigma$ is ReLU activation.  The spectral convolution $\mathcal{K}_l$ is defined as:
\begin{equation}
(\mathcal{K}_l v)(x)
= \mathcal{F}^{-1}\!\left[
R_l(k_n)\,\hat{v}(k_n)
\right](x),
\qquad
R_l(k_n)\in\mathbb{C}^{d_v\times d_v}.
\end{equation}
Here, we only keep the modes $n=0,\ldots,k_{\max}-1$, where $k_\mathrm{max} = 128$ modes, and all remaining Fourier coefficients are set to zero.

\paragraph{FNO training without and with redundancy}  We test two training schemes for the FNO, generating two surrogates as the result, $\mathcal{S}^\text{FNO-B}$ and $\mathcal{S}^\text{FNO-R}$.  For the first one, $\mathcal{S}^\text{FNO-B}$, we use only the basic state variables, $\ln V$ and $\Theta$:\footnote{We use $\ln V$ rather than $V$ because, at a given time, $V$ spans many orders of magnitude along the fault.}

\begin{equation}
    \mathcal{S}^{\text{FNO-B}}: (\ln V, \Theta) \mapsto (\Delta \ln V, \Delta \Theta),
\end{equation}
where $\Delta$ denotes the change in the variable from before to after the event.\footnote{We use these changes as the FNO outputs because we find that this gives better training performance.} For the second one, $\mathcal{S}^\text{FNO-R}$, we add redundancy by incorporating the values of the shear stress $\tau$ and the change in the functional $\Delta \phi$ into training:

\begin{equation}
    \mathcal{S}^{\text{FNO-R}}: (\ln V, \Theta,\tau) \mapsto (\Delta \ln V, \Delta \Theta, \Delta \phi).
\end{equation}

In principle, $\tau$ and $\Delta \phi$ can be computed from the values of $V$ and $\Theta$ before and after the jump, so the two jump formulations are theoretically identical.  However, we find that the second redundant formulation produces more accurate results, as discussed below.  Note that $\Delta \phi$ is closely related to the change in $\tau$; they differ only by the values of the radiation damping term $\frac{\mu}{2c_S} V$, which should be negligibly small before and after an earthquake event.

\paragraph{Loss function and the optimizer} To quantify the accuracy of the surrogates, we define the following loss function:
\begin{equation}\label{eq:FNOloss}
\mathcal{L} = \sum_{i}\frac{\|\hat{x_i} - x_i \|_2}{\| x_i \|_2}
\end{equation}
where $\| \cdot \|_2$ is the $L^2$ norm, $x$ is a collection of values after earthquake events from the conventional simulation, and $\hat{x}_i$ are computed from the values generated by the FNO surrogates. We define $x = \{\ln {V}, \Theta\}$ for $\mathcal{S}^\text{FNO-B}$ and $x = \{\ln {V}, \Theta, \Delta\phi, \Delta\delta, V, \phi[V]\}$ for $\mathcal{S}^\text{FNO-R}$, where we use the FNO output $\Delta \phi$ and the  quasi-dynamic approximation $ \widehat{\Delta \delta}(k)  = -\frac{2}{\mu|k|}\;\widehat{\Delta\phi}(k)$ to compute $\Delta \delta$.   Note that the addition of $\Delta\delta$, $V$, and $\phi[V]$ to the loss function improves the accuracy of the hybrid method below.






\subsection{Hybrid Simulator}
We propose a hybrid architecture to accelerate the earthquake sequence simulation, combining a conventional QD solver in the interseismic periods with the learned FNO jump operator for the earthquake (seismic) events. Outside the seismic event intervals, the QD approximation is accurate and the dynamics are local in time, so we evolve the system using QD until an event is detected. During each
event interval, we replace the expensive FD rupture dynamics by applying the learned jump
map $\mathcal{S}^\text{FNO}$.

Let $\mathcal{S}^{\mathrm{QD}}$ denote the QD evolution operator that advances the state from the end of event $i-1$ to the start of event $i$, i.e., from $t_e^{i-1}$ to $t_s^i$ (where $t_e^0$ corresponds to the initial time). Given an initial state $U_0$ (assumed to lie sufficiently close to the attracting set and outside the event set), the hybrid evolution after $n$ events is the composition:
\begin{equation}
    \mathcal{S}_n(U_0) : =  \underbrace{(\mathcal{S}^\text{FNO} \circ \mathcal{S}^{\mathrm{QD}}\circ \cdots \circ \mathcal{S}^\text{FNO} \circ \mathcal{S}^{\mathrm{QD}})}_{n \text{ times}}(U_0).
\end{equation}
Hence, we have the following algorithm.
\begin{mdframed}
{\bf Algorithm H}

\begin{enumerate}
  \item \textbf{Interseismic (QD) phase:} Start from the initial conditions or the state after the previous event, $U(\cdot,t_e^{i-1})$, integrate the QD system until the event-start
  criterion is satisfied (event detected at $t=t_s^i$, $i\in\{1,\dots , n\}$), and record the pre-event state $U(\cdot,t_s^i)$.
  \item \textbf{Seismic (FNO surrogate) phase:} Apply the learned jump operator with $U(\cdot,t_s^i)$ as input to obtain the
  post-event state, $U(\cdot,t_e^i)=\mathcal{S}^\text{FNO}\bigl(U(\cdot,t_s^i)\bigr)$.
  \item Repeat Steps 1--2 until the desired number of events is reached.
\end{enumerate}
\end{mdframed}

\section{Comparison between the conventional and hybrid simulations} \label{sec:results}
\subsection{Performance of the FNO map}

\begin{figure}
    \centering
    \includegraphics[width=1\linewidth]{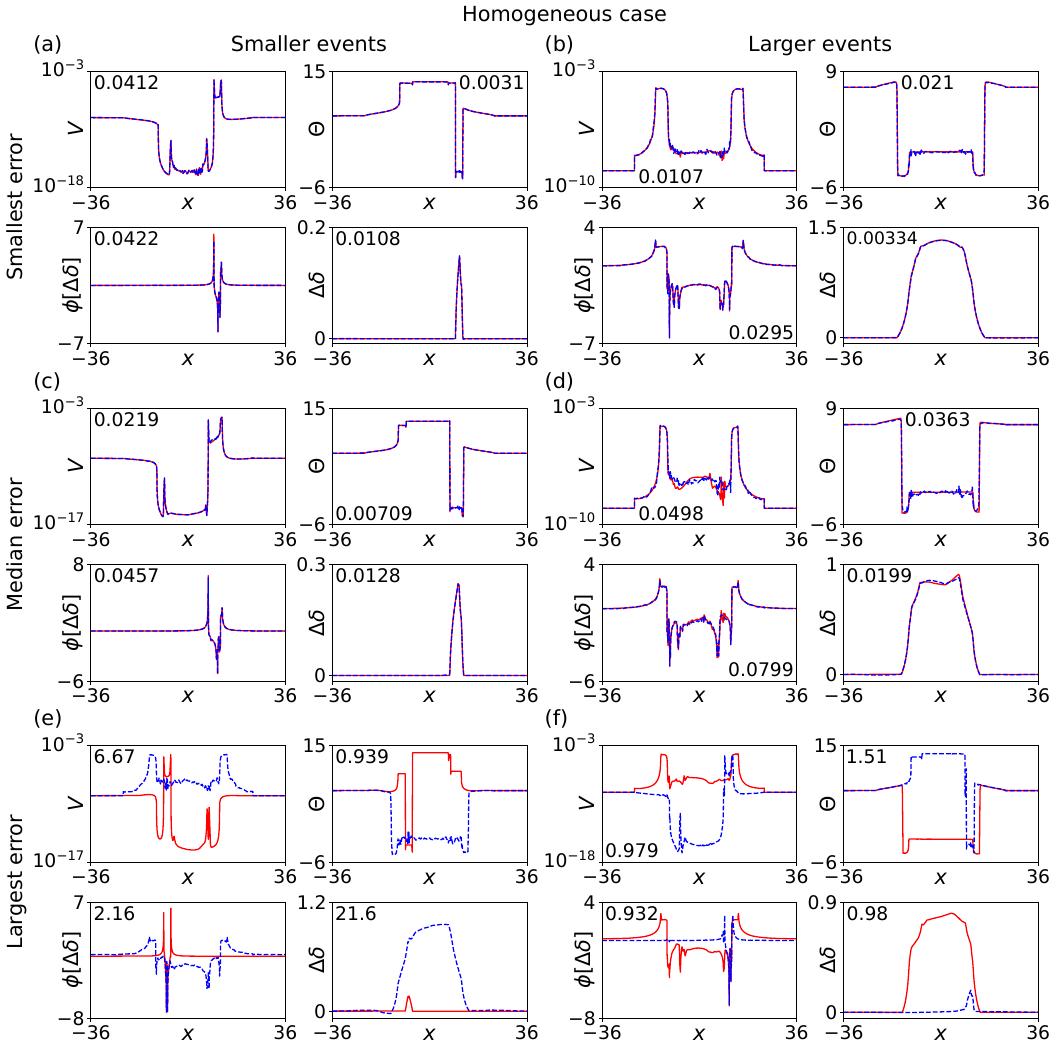}
    \caption{Comparison of fully dynamic (FD) solutions (red solid lines) and FNO jump-map predictions (blue dashed lines) for the homogeneous case. Panels (a),(c),(e) show smaller events, and (b),(d),(f) show larger events. Within each event class, examples are selected from the test set according to the smallest (a,b), median (c,d), and largest (e,f) mean error, defined as the average of the relative $L^2$ errors in the slip rate $V$ and state variable $\Theta$. The numbers in each plot indicate the relative $L^2$ error of the corresponding field.}
    \label{fig:homo}
\end{figure}

\begin{figure}
    \centering
    \includegraphics[width=1\linewidth]{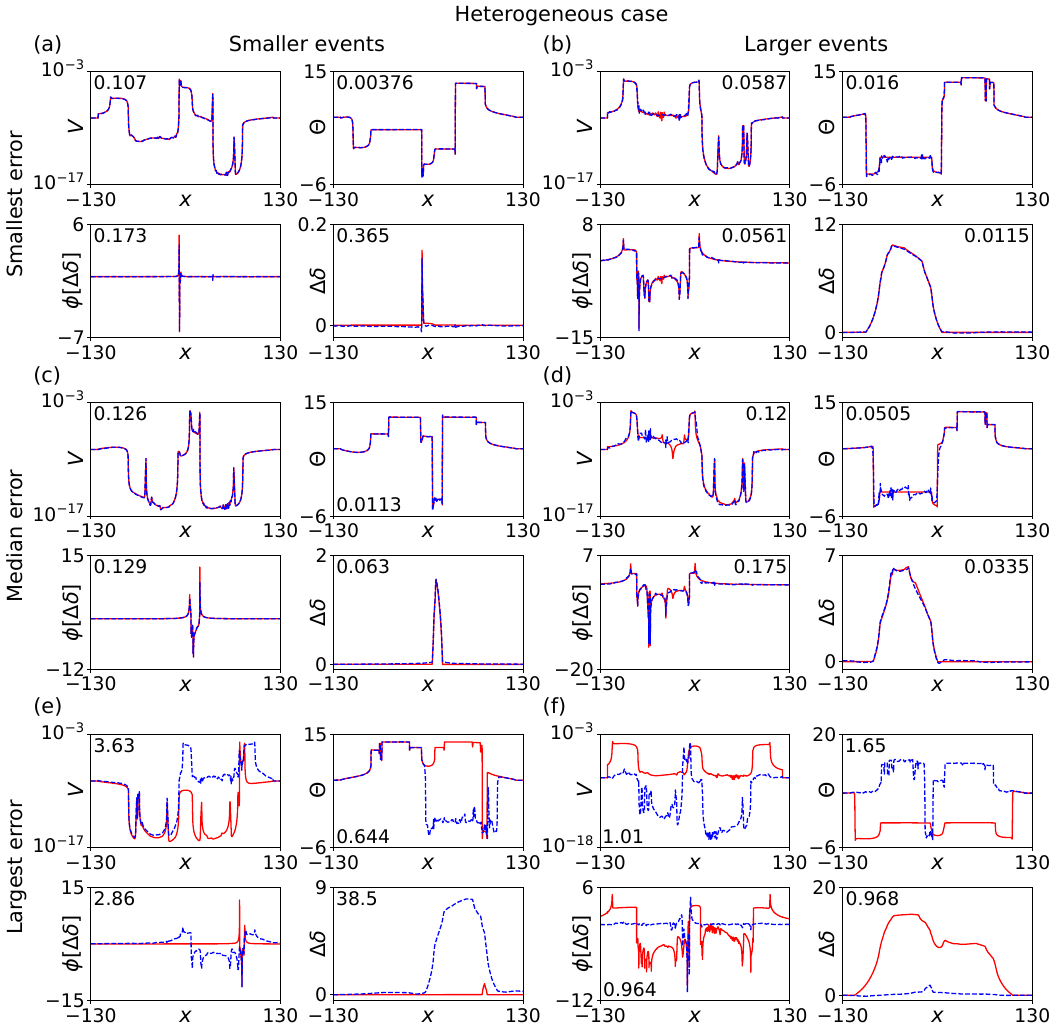}
    \caption{Comparison of fully dynamic (FD) reference solutions (red solid lines) and FNO jump-map predictions (blue dashed lines) for the heterogeneous case. Panels (a,c,e) show smaller events, and (b,d,f) show larger events. Within each event class, examples are selected from the test set according to the smallest (a,b), median (c,d), and largest (e,f) mean error, defined as the average of the relative $L^2$ errors in the slip rate $V$ and state variable $\Theta$. The numbers in each plot indicate the relative $L^2$ error of the corresponding field.}
    \label{fig:hetero}
\end{figure}

\begin{figure}
    \centering
    \includegraphics[width=1\linewidth]{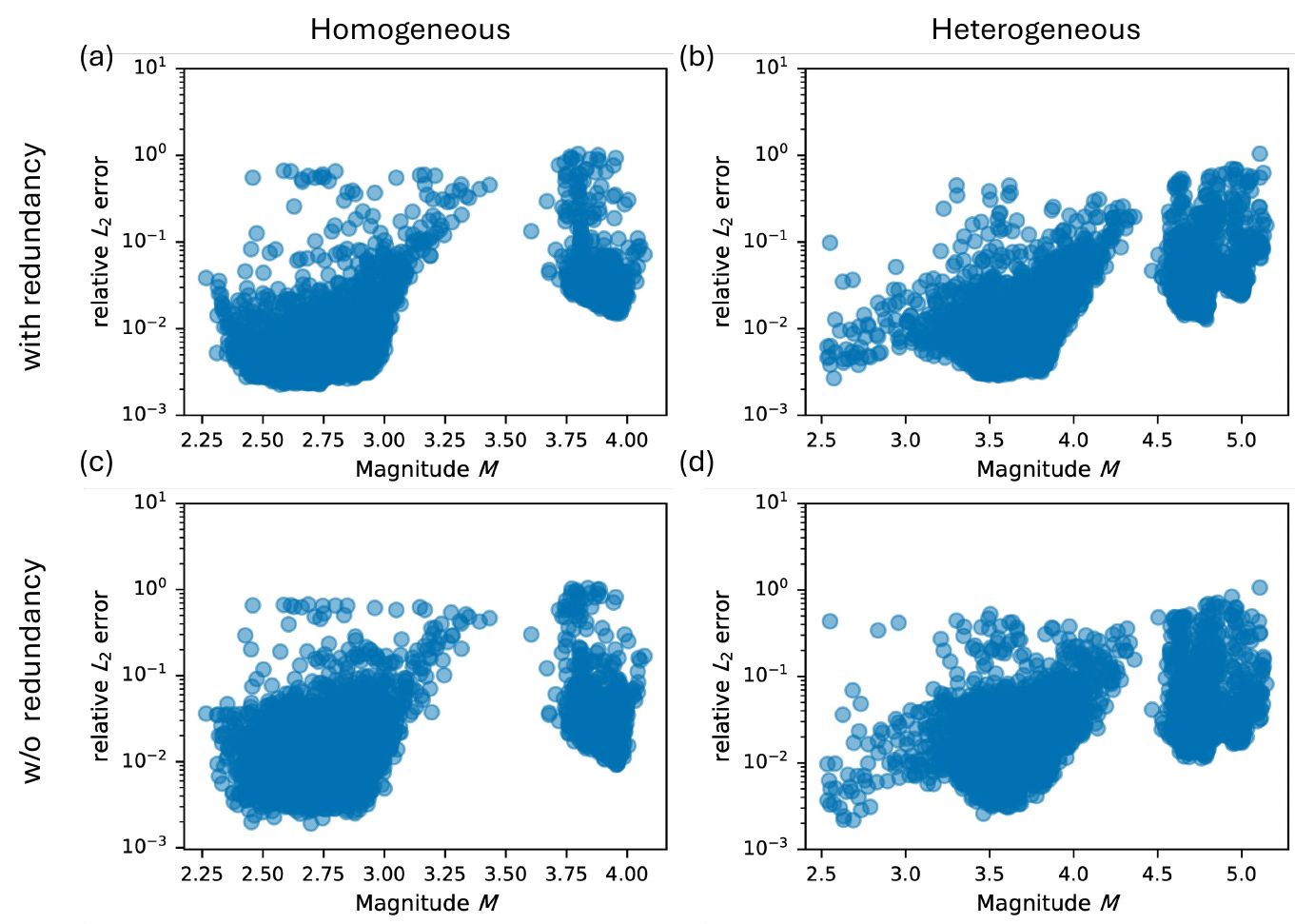}
    \caption{Relative $L^2$ error of the FNO map as a function of moment magnitude $M$ for the homogeneous (a,c) and heterogeneous (b,d) cases. Each point represents one test event, with the error defined as the average of the relative $L^2$ errors in slip rate $V$ and state variable $\Theta$. The top row shows $\mathcal{S}^\text{FNO-R}$ (trained with redundancy), and the bottom row shows $\mathcal{S}^\text{FNO-B}$ (trained without redundancy).}
    \label{fig:error}
\end{figure}

\begin{figure}
    \centering
    \includegraphics[width=1\linewidth]{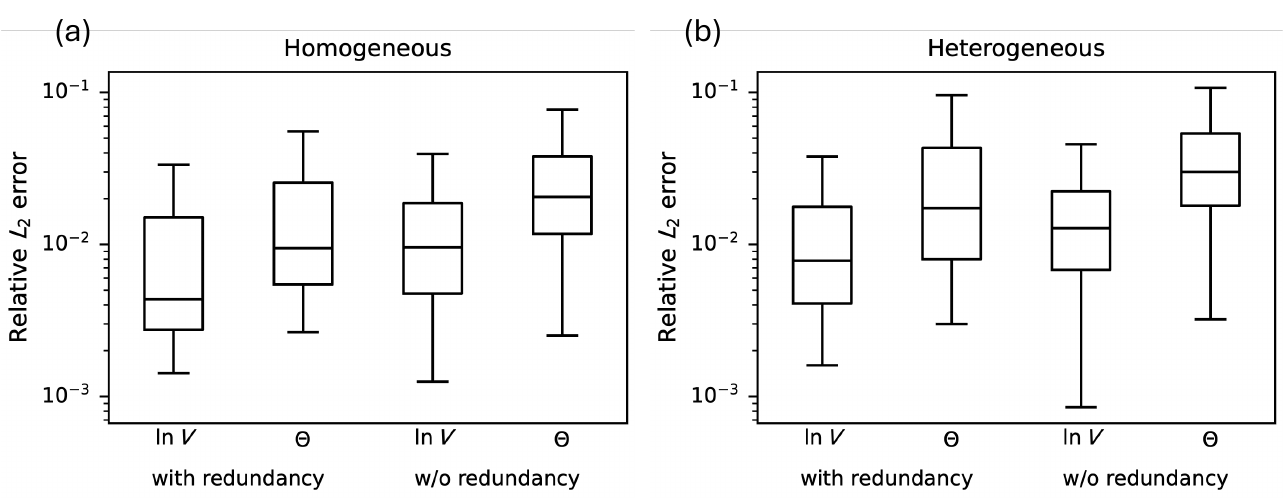}
    \caption{Relative $L^2$ errors in slip rate $\ln V$ and state variable $\Theta$ over the test set, comparing $\mathcal{S}^\text{FNO-R}$ and $\mathcal{S}^\text{FNO-B}$ (trained with and without redundancy) in the (a) homogeneous and (b) heterogeneous cases. Boxes show the interquartile range, and middle lines indicate the medians.}
    \label{fig:boxplot}
\end{figure}

We evaluate the FNO jump map on the test samples, comparing $\mathcal{S}^\text{FNO-R}$ and $\mathcal{S}^\text{FNO-B}$ trained with and without redundancy.  The overall error of the jump map predicted by the trained FNO compared to the test samples from the fully dynamic (FD) model is shown in Table~\ref{tab:fnomap_error}.  The median relative errors in the test dataset are generally below 0.01 for the logarithmic slip rate, 0.02 for the state variable, and  0.14 for the shear stress change in both the homogeneous and heterogeneous cases.  Figures~\ref{fig:homo} and \ref{fig:hetero} illustrate three cases representing the smallest, median, and largest error of $\mathcal{S}^\text{FNO-R}$.  The smallest-error and median-error examples closely reproduce the spatial profiles of slip rate, state, stress change, and coseismic slip.  Even for events with large errors, including $V$ in the loss function helps control the maximum predicted slip rate, which is important for stable continuation of the hybrid simulation.  The distribution of errors is shown in Figures \ref{fig:error} and \ref{fig:boxplot}.

Including redundancy in the mapping and loss function improves prediction accuracy for both homogeneous and heterogeneous cases, as shown by Table~\ref{tab:fnomap_error} and Figure~\ref{fig:boxplot}.  The median logarithmic slip-rate error decreases from 0.0172 to 0.0091 for the homogeneous case and from 0.0128 to 0.0078 for the heterogeneous case. The median state-variable error decreases from 0.0393 to 0.0207 for the homogeneous case and from 0.0300 to 0.0173 for the heterogeneous case.

The results so far (Table~\ref{tab:fnomap_error}) are based on training with the same spatial resolution as the conventional simulator output (8192), which we call ``original''. To test the invariance of the FNO surrogate with respect to the spatial discretization, we train the FNO model with redundancy on the datasets downsampled to 4096 and then test its performance with various discretizations, from the original one of 8192 down to 512 (Table~\ref{tab:disc-invariance}). We find that the FNO surrogate performs well at all resolutions tested, demonstrating the discretization invariance of the FNO surrogate, although the accuracy declines at the lowest resolution.

We conclude that the FNO can learn an accurate approximation of the non-Markovian earthquake jump map in both homogeneous and heterogeneous cases, with discretization invariance and with improved results when we include redundancy in the mapping and loss function.


\begin{table}[H]
    \caption{Median relative $L^2$ errors over the test samples for the homogeneous and heterogeneous cases}
    \label{tab:fnomap_error}
    \centering
    \begin{tabular}{l c c}
        Variable & $\mathcal{S}^\text{FNO-R}$ (with redundancy) & $\mathcal{S}^\text{FNO-B}$ (without redundancy)\\
        \midrule
        \midrule
        \multicolumn{3}{l}{\textit{Homogeneous}} \\
        $\ln V$ & 0.00436 & 0.00957\\
        $\Theta$ & 0.00944 & 0.02055\\
        $\phi[\Delta\delta]$ & 0.06883 & --\\
        \midrule
        \multicolumn{3}{l}{\textit{Heterogeneous}} \\
        $\ln V$ & 0.00782 & 0.01280\\
        $\Theta$ & 0.01731 & 0.03000\\
        $\phi[\Delta\delta]$ & 0.13393 & --\\
        \midrule
        \midrule
    \end{tabular}
\end{table}

\begin{table}\label{tab:disc-invariance}
\caption{Median relative $L^2$ errors for the FNO surrogate trained with down-sampled datasets and tested at different discretizations for 1200 homogeneous and 4000 heterogeneous test events.}
\centering
\begin{tabular}{l c c c c c}
\multicolumn{6}{l}{\textit{Homogeneous}} \\
Variable & 8192 (sampled) & 4096 (trained) & 2048 & 1024 & 512 \\
\midrule
\midrule
$\ln V$ & 0.00829 & 0.00847 & 0.01102 & 0.01126 & 0.02440 \\
$\Theta$ & 0.02022 & 0.01997 & 0.02610 & 0.02820 & 0.06505 \\
$\phi[\Delta\delta]$ & 0.13588 & 0.13439 & 0.15310 & 0.15326 & 0.21393 \\
\midrule
\multicolumn{6}{l}{\textit{Heterogeneous}} \\
Variable & 16384 (sampled) & 8192 (trained) & 4096 & 2048 & 1024 \\
\midrule
\midrule
$\ln V$ & 0.00748 & 0.00711 & 0.00844 & 0.01356 & 0.02480 \\
$\Theta$ & 0.01890 & 0.01761 & 0.02306 & 0.03532 & 0.06362 \\
$\phi[\Delta\delta]$ & 0.13093 & 0.12497 & 0.15911 & 0.19191 & 0.34910 \\
\midrule
\midrule
\end{tabular}
\end{table}


\subsection{Hybrid Simulator}
We evaluate the hybrid simulator by comparing its long-term earthquake statistics with those of the fully dynamic (FD) reference simulation and its quasi-dynamic (QD) approximation. The simulations span 20,000 years for the homogeneous case and 40,000 years for the heterogeneous case.

\begin{figure}[t]
    \centering
    \includegraphics[width=1\linewidth]{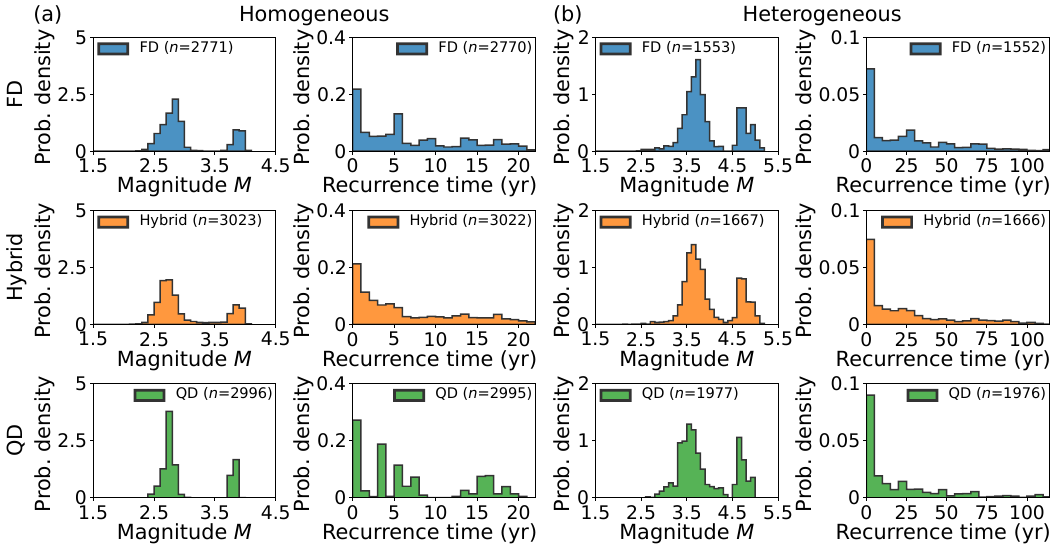}
    \caption{Probability distributions of event moment magnitude $M$ and recurrence time in the (a) homogeneous and (b) heterogeneous cases. Rows show fully dynamic (FD), hybrid, and quasi-dynamic (QD) simulations, respectively. Within each case, the left column shows magnitudes and the right column shows recurrence times. Simulations span 20,000 years in the homogeneous case and 40,000 years in the heterogeneous case. The number of events or recurrence intervals is indicated by $n$.}
    \label{fig:hybridresult}
\end{figure}

\paragraph{Performance}
We begin by examining the distributions of event magnitudes and recurrence times. Figure~\ref{fig:hybridresult} compares the distributions obtained from the FD simulations (first row) with those obtained from the hybrid simulation (second row).  The quantitative comparison is shown in Table~\ref{tab:hybrid} using both the Wasserstein distance and KL divergence.  We see good agreement.    This figure and table also compare the FD simulations (first row) with the state-of-the-art approximation using QD simulations (third row).  We notice that the performance of the hybrid simulation is substantially better than that of the widely used QD approximation. This is because QD does not retain any information about the inertial effects, while the hybrid simulation does, through training on the fully dynamic data.


\begin{table}
\caption{Wasserstein distances and KL divergences between event catalogs}
\label{tab:hybrid}
\centering
\begin{tabular}{l c c c c c c}
comparison
& $W_2(M)$
& $\mathrm{KL}(M)$
& $W_2(\mathrm{rec})$ [yr]
& $\mathrm{KL}(\mathrm{rec})$ \\
\midrule
\midrule
\multicolumn{5}{l}{\textit{Homogeneous}} \\
FD vs Hybrid & 0.0906 & 0.1517 & 0.7899 & 0.0804 \\
FD vs QD     & 0.1778 & 0.6468 & 1.1200 & 0.6866 \\
\midrule
\multicolumn{5}{l}{\textit{Heterogeneous}} \\
FD vs Hybrid & 0.0487 & 0.0381 & 3.4178 & 0.0422 \\
FD vs QD     & 0.0959 & 0.1091 & 7.5223 & 0.1334 \\
\midrule
\midrule
\end{tabular}
\end{table}

\begin{figure}
    \centering
    \includegraphics[width=1\linewidth]{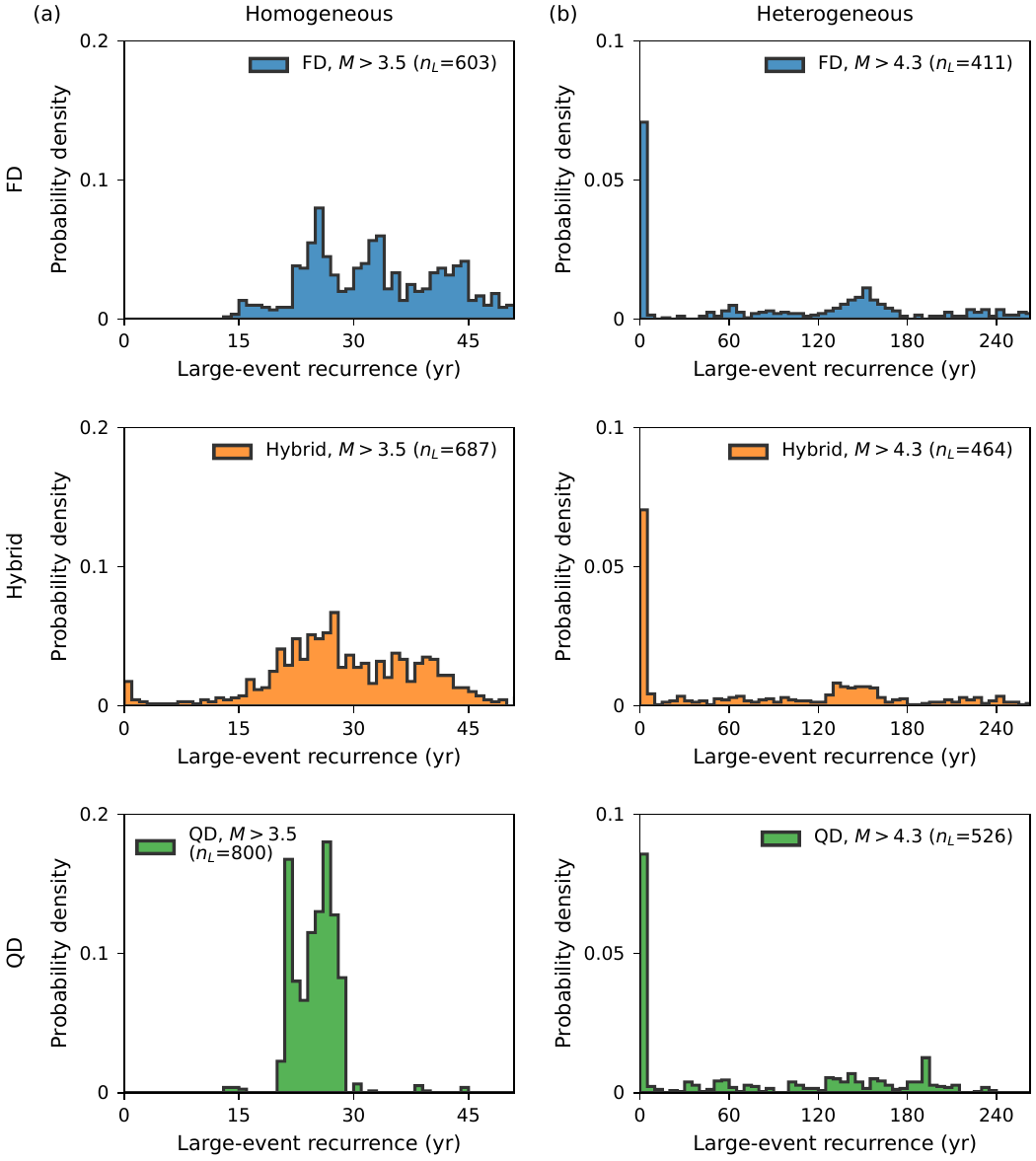}
    \caption{Probability distributions of recurrence times between successive larger events in the (a) homogeneous and (b) heterogeneous cases. Larger events are defined by magnitudes $M>3.5$ and $M>4.3$, respectively. Rows show fully dynamic (FD), hybrid, and quasi-dynamic (QD) simulations. Recurrence times are measured between the start times of successive larger events, and $n_L$ denotes the number of the larger events.}
    \label{fig:hybrid_recurrence}
\end{figure}

We also examine higher-order statistics by considering the recurrence times between larger events only and transition probabilities between smaller and larger events (Figure~\ref{fig:hybrid_recurrence} and Tables~\ref{tab:hybrid_recurrence} and~\ref{tab:markov_stats}). Larger events are defined by magnitude thresholds of 3.5 and 4.3 for the homogeneous and heterogeneous cases, respectively, motivated by the bimodal magnitude distributions.   We find that the hybrid simulations agree well with the FD simulations in both magnitude and transition probability and perform much better than the QD simulations. 

We conclude that the proposed hybrid approach reproduces the statistical features of the FD earthquake magnitude and recurrence time well, and more accurately than the state-of-the-art QD approximation. 


\begin{table}
\caption{Wasserstein distances and KL divergences between recurrence time distributions of larger events}
\label{tab:hybrid_recurrence}
\centering
\begin{tabular}{l c c c}
comparison & $W_2$ [yr] & $\mathrm{KL}$ \\
\midrule
\midrule
\multicolumn{3}{l}{\textit{Homogeneous}} \\
FD vs Hybrid & 4.5887  & 0.2439 \\
FD vs QD     & 10.3293 & 0.1910 \\
\midrule
\multicolumn{3}{l}{\textit{Heterogeneous}} \\
FD vs Hybrid & 15.3687 & 0.1434 \\
FD vs QD     & 31.0145 & 0.1909 \\
\midrule
\midrule
\end{tabular}
\end{table}

\begin{table}
\caption{Transition probabilities between smaller (S) and larger (L) events}
\label{tab:markov_stats}
\centering
\begin{tabular}{l c c c c c c}
catalog & $P(S|S)$ & $P(L|S)$ & $P(S|L)$ & $P(L|L)$ & $n(S)$ & $n(L)$\\
\midrule
\midrule
\multicolumn{7}{l}{\textit{Homogeneous} (threshold $M > 3.5$)}\\
FD & 0.722 & 0.278 & 0.998 & 0.002 & 2167 & 603\\
Hybrid & 0.713 & 0.287 & 0.975 & 0.025 & 2335 & 687\\
QD & 0.636 & 0.364 & 1.000 & 0.000 & 2196 & 799\\
\midrule
\multicolumn{7}{l}{\textit{Heterogeneous} (threshold $M > 4.3$)}\\
FD & 0.754 & 0.246 & 0.684 & 0.316 & 1141 & 411\\
Hybrid & 0.741 & 0.259 & 0.670 & 0.330 & 1202 & 464\\
QD & 0.780 & 0.220 & 0.606 & 0.394 & 1450 & 526\\
\midrule
\midrule
\end{tabular}
\end{table}

\paragraph{Computational effort}

Table~\ref{tab:hybrid_cost} shows the computational cost for a 20,000-year homogeneous benchmark and a 40,000-year heterogeneous benchmark using 16 processes. In the homogeneous case, the total computation time decreases from 84.0 hours for FD to 11.1 hours for the hybrid model providing an approximately 7.6-fold decrease in computational time. This is because the average time to simulate an earthquake event in FD is 88.3 s compared to a meager 0.0217 s to evaluate the FNO in the hybrid simulation. Further, the hybrid approach eliminates the 1.61 GB memory required to store the history in the FD simulation.

In the heterogeneous case, the total computation time decreases from 136.9 hours for FD to 11.2 hours for the hybrid model providing an approximately 12.2-fold decrease in computational time. The average time to simulate an earthquake event in FD is 242 s compared to 0.0325 s to evaluate the FNO in the hybrid simulation. Further, the hybrid model eliminates the 4.29 GB memory required to store the history in the FD simulation.

The table also shows that the hybrid model requires less computational effort than the QD simulations by a factor of approximately 1.35 in the homogeneous case and 1.71 in the heterogeneous case, even though it produces a better approximation.


\begin{table}
    \caption{Computational cost with 16 processes for the homogeneous and heterogeneous cases.}
    \label{tab:hybrid_cost}
    \centering
    \begin{tabular}{l c c c}
        & \shortstack{Entire simulation}
        & \shortstack{Average time\\per earthquake}
        & \shortstack{Memory to store\\the history} \\
        \midrule
        \midrule
        \multicolumn{4}{l}{\textit{Homogeneous} (20,000-year simulation)} \\
        FD & 84.0 h & 88.3 s & 1.61 GB \\
        FNO hybrid & 11.1 h & 0.0217 s & 0 \\
        QD & 15.0 h & 10.8 s & 0 \\
        \midrule
        \multicolumn{4}{l}{\textit{Heterogeneous} (40,000-year simulation)} \\
        FD & 136.9 h & 242 s & 4.29 GB \\
        FNO hybrid & 11.2 h & 0.0325 s & 0 \\
        QD & 19.2 h & 22.1 s & 0 \\
        \midrule
        \midrule
    \end{tabular}
\end{table}

\section{Conclusion}\label{sec:conclusion}
We develop a hybrid framework that uses a Fourier neural operator (FNO) to accelerate fully dynamic earthquake sequence simulations. The framework addresses the multiscale, chaotic, and history-dependent dynamics of earthquakes by combining a learned earthquake jump map with a conventional quasi-dynamic solver. The FNO maps the fault state immediately before an earthquake to the state immediately after it. Augmenting its input and output spaces with redundant variables improves the accuracy of the FNO surrogate. In our hybrid simulation scheme, the quasi-dynamic solver advances the slow interseismic evolution, during which the dynamic contribution to stress transfer is negligible; the learned jump map captures the effects of fully dynamic rupture.

We generate training data efficiently from an ensemble of short, independent simulations run in parallel. After initial transients are discarded, the ensemble yields magnitude and recurrence-time distributions that closely agree with those of a single long simulation; this agreement supports using the ensemble to efficiently sample representative earthquake events.

For both homogeneous and heterogeneous fault cases, the hybrid simulations reproduce the magnitude and recurrence-time distributions of earthquake events from the fully dynamic reference simulation more accurately than the commonly used quasi-dynamic approximation. The hybrid simulations also accurately capture higher-order statistics, such as the recurrence time of larger events and transitions between smaller and larger events. Once trained, the hybrid model runs about 7.6 and 12.2 times faster than the fully dynamic solver in the 20,000-year homogeneous and 40,000-year heterogeneous benchmarks, respectively, and about 1.4 and 1.7 times faster than the quasi-dynamic solver. It eliminates the need to store the slip-rate history. These results demonstrate that learning earthquake jump maps provides an efficient approach to long-term sequence simulation while retaining the influence of dynamic rupture on earthquake statistics.

This work has several promising extensions. First, our approach extends straightforwardly to the three-dimensional fault setting \cite{lapusta2009three_dimensional}. Second, the hybrid approach can be extended to models in which the interseismic periods include viscoplastic effects~\cite{allison2018earthquake, fang2024dynamic}, while the dynamic rupture can still be approximated by single-fault elastodynamics, as in this work. Third, the FNO can learn jump maps for faults with more complicated dynamic weakening mechanisms, such as flash heating and thermal pressurization of pore fluids~\cite{rice2006heating, thomas_quasidynamic_2014, lambert2021propagation}. The approach can also be extended to include variable friction properties of the fault.  



\section{Acknowledgments}
The authors thank Zongyi Li and Harkirat Singh for the FNO implementation used in this work. RY is grateful to the Funai Overseas Scholarship and the Japan Student Services Organization (JASSO) for their support, KB to the Office of Naval Research through N00014-23-1-2654, and NL to the Center for Geomechanics and Mitigation of Geohazards (GMG) at Caltech.

\section{AI Usage}
The authors used Claude Opus 5.5 and earlier Claude models (Anthropic) to revise the manuscript and code. The authors reviewed all changes and take full responsibility for this article.

\bibliographystyle{abbrv}
\bibliography{references}

\newpage
\appendix
\section{Additional discussion}
\label{sec:technical}

\subsection{BICyclE (Boundary Integral for Cycles of Earthquakes) Algorithm}
Here, we review the method of~\cite{lapusta2000}, which we use to simulate the dynamical system in \eqref{eq:FD}.
\paragraph{Truncation of the wave-mediated stress effects}
The wave-mediated stress effects are only important during earthquake events and negligible during long interseismic periods.  Hence we compute the convolution in \eqref{eq:phi} only within the most recent time window $T_w$  chosen such that the contribution to the functional $\phi$ of the deformation history occurring prior to time $t-T_w$ is negligible:
\begin{equation}\label{eq:trancated phihat}
\widehat{\phi}(k,t)
= -\frac{\mu|k|}{2}\left[
\widehat{\delta}(k,t)
-\int_{0}^{T_w}
W\!\bigl(|k|c_s t'\bigr)\,\widehat{V}\!\bigl(k,(t)-t'\bigr)\,dt'
\right].
\end{equation}
We use the same window $T_w$ for all Fourier modes (the method of~\cite{lapusta2000} also allows mode-dependent windows) equal to the time for the shear wave to propagate through the VW, seismogenic domain, which gives the approximate duration of the largest earthquake events in the model.
\subsubsection{Updating scheme}

\subsubsection*{Algorithm}
Suppose $\delta(x, t), V(x, t), \Theta(x,t)$, and the history of $\{V(x,t'), t'\in [t-T_w, t]\}$  are known; we compute $\delta(x, t+\Delta t), V(x, t+\Delta t), \Theta(x,t+\Delta t)$ as follows. Let
\[
\mathcal{E}_s(\Theta;V):=\ln\Bigl(\frac{V^*}{V}+\Bigl(e^{\Theta}-\frac{V^*}{V}\Bigr)e^{-Vs/D_{\rm RS}}\Bigr)
\]
denote the state that the aging law \eqref{eq:aging} reaches from $\Theta$ after a time $s$ with $V$ held fixed, and mark predicted values with a tilde.
\begin{enumerate}
  \item Compute the first prediction of  $\tilde\delta(x, t+\Delta t)$, $\tilde\Theta(x,t+\Delta t)$, and $\tilde\phi(x, t+\Delta t)$ by assuming that the slip rate is constant through the time step at $V(x,t)$.
  \begin{align*}
      &\tilde\delta(x, t+\Delta t) = \delta(x,t) + \Delta t V(x,t)\\
      &\tilde\Theta(x,t+\Delta t) = \mathcal{E}_{\Delta t}\bigl(\Theta(x,t);V(x,t)\bigr)
  \end{align*}
  \begin{equation}\label{eq:trancated phihat}
\widehat{\phi}(k,t+\Delta t)
= -\frac{\mu|k|}{2}\left[
\widehat{\delta}(k,t+\Delta t)
-\underbrace{\int_{\Delta t}^{T_w}
W\!\bigl(|k|c_s t'\bigr)\,\widehat{V}\!\bigl(k,(t+\Delta t)-t'\bigr)\,dt'}_{\text{depends on stored history from } t+\Delta t-T_w \text{ up to } t}
-\underbrace{\widehat{V}(k,t)\int_{0}^{\Delta t} W\!\bigl(|k|c_s t'\bigr)\,dt'}_{\text{current-step approximation}}
\right].
\end{equation}
  
  \item Compute the updated $\tilde V(x, t + \Delta t)$ and $\tau(x,t+\Delta t)$ using \eqref{eq:bie} and \eqref{eq:rs} by Newton--Raphson search.
  
  \item Correct the slip, state variable, and functional using the updated $\tilde V$, integrating the state over the first half of the step at $V(x,t)$ and over the second half at $\tilde V(x,t+\Delta t)$,
    \begin{align*}
      &\delta(x, t+\Delta t) = \delta(x,t) + \Delta t \left( \frac{V(x,t)+\tilde V(x, t + \Delta t)}{2}\right )\\
      &\Theta(x,t+\Delta t) = \mathcal{E}_{\Delta t/2}\Bigl(\mathcal{E}_{\Delta t/2}\bigl(\Theta(x,t);V(x,t)\bigr);\,\tilde V(x,t+\Delta t)\Bigr)
    \end{align*}

  \item Repeat Step 2 with $\delta(x,t+\Delta t)$ and $\Theta(x,t+\Delta t)$, using $\bigl(\widehat{V}(k,t)+\widehat{\tilde V}(k,t+\Delta t)\bigr)/2$ in place of $\widehat{V}(k,t)$ in the current-step term, to obtain $V(x,t+\Delta t)$. Store $\bigl(\widehat{V}(k,t)+\widehat{V}(k,t+\Delta t)\bigr)/2$ as the history on $[t,t+\Delta t]$.
\end{enumerate}

\paragraph{Time step}
After each step, we choose the next time step as
\[
\Delta t_{\mathrm{new}}=\min\Bigl\{2\Delta t,\ \max\bigl\{\Delta t_{\min},\ \min\bigl\{\Delta t_{\max},\ \min_{x\in\Gamma_f}\zeta(x)D_{\rm RS}/V(x,t+\Delta t)\bigr\}\bigr\}\Bigr\},
\]
and, if $\Delta t_{\mathrm{new}}\leq T_w$, we round it down to a multiple of $\Delta t_{\min}$, the spacing of the stored history. Here $\zeta=\min\{\zeta_L/2,\ 0.33\}$, where $\zeta_L$ is the expression in Eq.~(15) of~\cite{lapusta2000}, without its cap of $1/2$, evaluated with the single-cell stiffness $\pi\mu/(4\Delta x)$; $\Delta t_{\min}=\Delta x/(\beta c_s)$; and the cap $\Delta t_{\max}$ is about 2.5 days in the homogeneous case and 0.2\,yr in the heterogeneous case.

\subsection{When does the QD approximation hold?}

\begin{table}[H]\label{QDthreshold}
\centering
\caption{
Maximum dynamic contribution in the QD regions.
Each value is the maximum of
$\|\phi_{\mathrm{dyn}}\|_p/\|\phi\|_p$. Values are the worst cases over all relevant
time steps in events 11--30}.
\label{tab:qd-validity}
\begin{tabular}{lllccc}
\toprule
Model & Period & QD condition & $L_1$ & $L_2$ & $L_\infty$ \\
\midrule
Homogeneous
& Before event
& $\|V\|_\infty<10^{-3}\,\mathrm{m/s}$
& $9.7\times10^{-5}$
& $3.5\times10^{-4}$
& $1.3\times10^{-3}$ \\

Homogeneous
& After event
& $\|V\|_\infty<10^{-4}\,\mathrm{m/s}$
& $5.0\times10^{-5}$
& $6.7\times10^{-5}$
& $1.2\times10^{-4}$ \\

Heterogeneous
& Before event
& $\|V\|_\infty<10^{-3}\,\mathrm{m/s}$
& $2.0\times10^{-5}$
& $1.0\times10^{-4}$
& $6.2\times10^{-4}$ \\

Heterogeneous
& After event
& $\|V\|_\infty<10^{-4}\,\mathrm{m/s}$
& $1.6\times10^{-5}$
& $2.0\times10^{-5}$
& $6.7\times10^{-5}$ \\
\bottomrule
\end{tabular}
\end{table}

\subsection{Reproducing the statistics of one long simulation with several shorter ones}\label{sec:initialcondition-dependence}
To produce the training data for the FNO surrogate, it is advantageous to run an ensemble of shorter simulations instead of one long simulation, since this can be done in parallel.  However, each simulation would have several initial events that are dominated by randomly selected initial conditions and are not representative of the system dynamics.  Here we compare the results of the ensemble of short simulations and one long simulation if we do not skip any events, skip the first 5 events, and skip the first 10 events (Figure 10). 
For the magnitude distributions, the 2-Wasserstein distances between the ensemble and the long run are 0.0945, 0.0510, and 0.0188, and the KL divergences are 0.1028, 0.0381, and 0.0060, if we skip no events, the first 5 events, and the first 10 events, respectively.
\begin{figure}[H]
    \centering
    \includegraphics[width=0.9\linewidth]{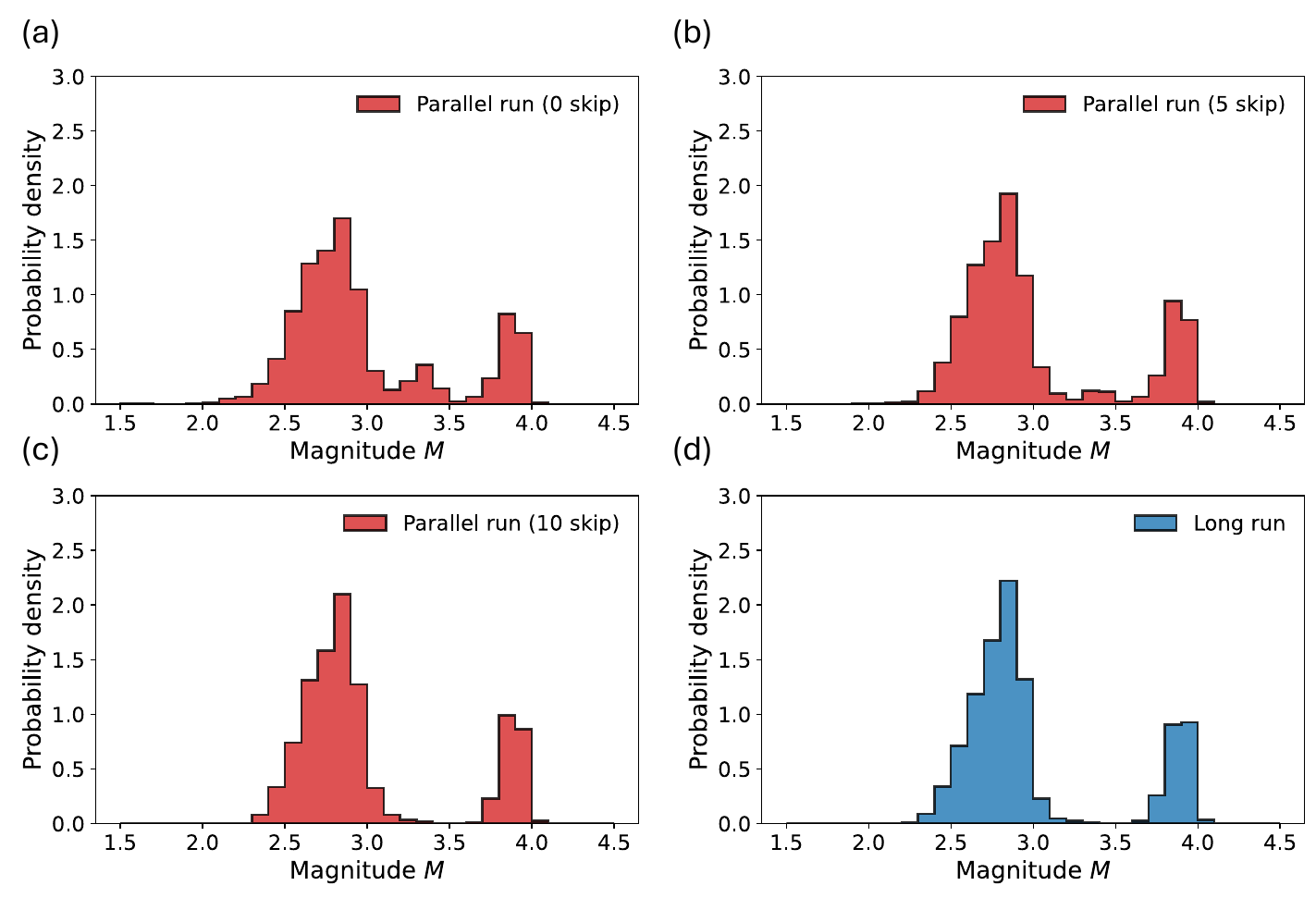}
    \caption{ Event distribution for an ensemble of simulations in which we (a) include all events,  (b) skip the first 5 events of each simulation, (c) skip the first 10 events of each simulation. (d) Event distribution for a single long run.}
    \label{fig:a-1}
\end{figure}

\subsection{Randomized initial conditions}\label{appendix: random-initial-condition}
 One way to choose initial conditions is to prescribe sliding at the plate rate $V=V_{\mathrm{pl}}$ everywhere on the fault, and the corresponding steady-state values of the state variable and shear stress. For each short run, we randomize this reference initial condition by adding five Gaussians to the initial value of the state variable,
\begin{equation}
\Theta(x,0)=\ln\frac{V^*}{V_{\mathrm{pl}}}+\sum_{j=1}^{5}A_j\exp\Bigl(-\frac{(x-x_j)^2}{2s_j^2}\Bigr),\qquad x\in\Gamma_f,
\end{equation}
where $\ln(V^*/V_{\mathrm{pl}})$ is the steady-state value at $V=V_{\mathrm{pl}}$. For each run, we draw the amplitudes $A_j$, centers $x_j$, and widths $s_j$ independently from uniform distributions: in the homogeneous case, $A_j\in[-4,4]$, $s_j\in[1,2]$ km, and $x_j$ lies in $\Gamma_{\mathrm{VW}}$; in the heterogeneous case, $A_j\in[-1,1]$, $s_j\in[2,5]$ km, and $x_j$ lies in the 160 km span of the two VW segments and the VS patch between them. Because we keep the initial stress $\tau_0$ of the reference state, the initial slip rate follows from equating \eqref{eq:bie} and \eqref{eq:rs} with the perturbed state variable.


\newpage

\section{Spring-slider system}
\label{sec:sprint-slider}
In this section, we study the quasi-static spring-slider system \cite{gu1984slip} as a finite-dimensional analog of our hybrid method. The spring-slider system consists of a slider connected to a spring of stiffness $k$ whose far end is pulled at a constant speed $V_0$. The frictional resistance of the slider is governed by rate-and-state friction, and the system can be written as an ordinary differential equation (ODE).

Although this model is simple, its dynamics are complex. For some parameter values, it shows stick-slip periodic cycles or chaotic sequences. With a single state variable, the quasi-static spring-slider is a two-dimensional autonomous system and therefore, by the Poincar\'e--Bendixson theorem, cannot be chaotic. With two state variables, the phase space is three-dimensional; as $k$ decreases below the critical stiffness $k = k_{\mathrm{cr}}$, the oscillations undergo period doubling and become apparently chaotic \cite{gu1984slip}. \cite{ShkollerMinster1997, Becker2000} confirmed deterministic chaos in this system by finding a positive Lyapunov exponent and approximately reduced its attractor to one-dimensional unimodal maps. A single slider with one state variable becomes chaotic only with inertia (the system is three-dimensional), and \cite{Erickson2008} found such chaos only for a strong velocity-weakening setting, beyond a period-doubling cascade that ends at $(b-a)/a \approx 11.5$. Spatially extended models can be chaotic at much weaker velocity weakening \cite{Erickson2011, Kato2014}.

Here, we consider the system with two state variables and the parameter setting record $J$ of~\cite{gu1984slip}, which produces the chaotic sequences. The parameter values are summarized in Table~\ref{tab:spring_slider_params}.

\subsection{Dynamical system equations}
Let $u$ be the displacement of the slider relative to the load point, with $u =0$ at steady sliding at $V_0$, and let $\Theta_1$ and $\Theta_2$ be the two state variables, which have units of stress ($\theta_1$, $\theta_2$ in \cite{gu1984slip}). The quasi-static force balance is
\begin{equation}\label{eqb:-stressbalance}
    \tau = \sigma_n f^* - k u = F(V,\Theta_1,\Theta_2),
\end{equation}
where $F$ is the friction law,
\begin{equation}\label{eqb:friction}
    F(V,\Theta_1,\Theta_2) = \sigma_n f^* + A \ln \frac{V}{V_0} + \Theta_1 + \Theta_2
\end{equation}
with the evolution of the state variables governed by the slip law \cite{ruina1983} as in Gu et al. \cite{gu1984slip}:
\begin{equation}
    \partial_t \Theta_i = G_i(V,\Theta_i) = -\frac{V}{L_i} \left( \Theta_i + B_i \ln{\frac{V}{V_0}} \right), \qquad i = 1,2,
\end{equation}
where $A = a\sigma_n$, $B_i = b_i \sigma_n$, and $V_0$ also serves as the reference slip velocity $V^*$. For the system state $x := (u,\Theta_1,\Theta_2) \in \mathbb{R}^3$, the dynamical system is the three-dimensional autonomous ODE $\dot{x} = f(x)$,

\begin{equation}\label{ode}
\dot{x} = f(x), \qquad f(x) = 
    \begin{pmatrix}
    V(x) -v_0 \\
    -\frac{V}{L_1} \left( \Theta_1 + B_1 \ln{\frac{V}{V_0}} \right) \\
    -\frac{V}{L_2} \left( \Theta_2 + B_2 \ln{\frac{V}{V_0}} \right)
    \end{pmatrix}
    ,
\end{equation}
where $V(x)$ is a function of the state $x$ computed from \eqref{eqb:-stressbalance}. Compared with \eqref{eq:FD}, the elastic interaction is through a single spring, and there is no radiation damping (which is mathematically equivalent to viscosity) and no history dependence (so the system evolution is Markovian). We integrate \eqref{ode} with an adaptive Runge--Kutta method (RK45).

\begin{table}[h]
\centering
\caption{Parameters of the spring-slider system (record J of \cite{gu1984slip}).}
\label{tab:spring_slider_params}
\begin{tabular}{lcc}
\toprule
Parameter & Symbol & Value \\
\midrule
Effective normal stress              & $\sigma_n$      & 3~MPa \\
Reference friction coefficient       & $f^*$           & 0.6 \\
Rate-and-state direct effect         & $a$             & 0.011 \\
Rate-and-state evolution effects     & $b_1,\ b_2$     & 0.011,\ 0.00924 \\
Characteristic lengths               & $L_1,\ L_2$     & 0.25~$\mu$m,\ $L_1/0.048 = 5.21~\mu$m \\
Load-point speed                     & $V_0$           & 1~$\mu$m/s \\
Spring stiffness                     & $k$             & $0.852\,k_{\mathrm{cr}} = 9.03 \times 10^{9}$~Pa/m \\
Event thresholds                     & $V_1 = V_2$     & $2V_0$ \\
\bottomrule
\end{tabular}
\end{table}

\subsection{Poincar\'e section and the event map}
Similar to \eqref{eq:event-start} and \eqref{eq:event-end}, we define the events using a velocity threshold and, for the case of the ODE, the threshold defines the Poincar\'e section,
\begin{equation}
    \Sigma := \{x \in \mathbb{R}^3: V = 2V_0\}.
\end{equation}
This is a plane in $\mathbb{R}^3$, which we split into the event start and end parts,
\begin{equation}
    \Sigma_s := \{ x \in \Sigma: \nabla V(x) \cdot f(x) > 0\}, \qquad \Sigma_e := \{ x \in \Sigma: \nabla V(x) \cdot f(x) < 0\}
\end{equation}
The fast and slow phases of each cycle define two Poincar\'e maps,
\begin{equation}
    S^{\dagger}: x(t_s) \in \Sigma_s \mapsto x(t_e) \in \Sigma_e, \qquad S^{\mathrm{load}}: x(t_e) \in \Sigma_e \mapsto x(t_s) \in \Sigma_s
\end{equation}

Analogous to the event magnitude and the recurrence time, we define the two measures, the event slip and the loading duration.

\begin{equation}
    \Delta u_n := u(t^n_e) - u(t^n_s), \qquad T^{\mathrm{load}}_n := t_s^n - t_e^{n-1}.
\end{equation}

\subsection{Learning the event map}

\paragraph{Dataset}
As in Section \ref{section:FNO}, we generate training data from an ensemble of short simulations. We run 20 simulations of 5,000 s, adding to $\Theta_1(0)$ a perturbation drawn uniformly from $[-55,-54)$ Pa. We ignore the first 625 s as a transient period, and sample the first 300 events of each run. We split the 6,000 samples into 4,200 for training, 600 for validation, and 1,200 for testing.

\paragraph{Feedforward Neural Network}
Since the state is finite-dimensional, we replace the FNO with a fully connected feedforward neural network. Let $\tilde a \in \mathbb{R}^3$ be the normalized input state $x(t^n_s)$ and $\tilde{y}\in \mathbb{R}^3$ the normalized output state $x(t^n_e)$, where each component is standardized with the mean and standard deviation of the training set. We use
\begin{equation}
\begin{aligned}
v_0 &= \sigma(P\tilde{a}), && P : \mathbb{R}^{3} \to \mathbb{R}^{d_v}\\
v_{l+1} &= \sigma(W_l v_l), && W_l: \mathbb{R}^{d_v} \to \mathbb{R}^{d_v}, \quad l = 0,\dots, 3, \\
\tilde{y}^{\mathrm{NN}} &= Qv_3,
&& Q: \mathbb{R}^{d_v} \to \mathbb{R}^3,
\end{aligned}
\end{equation}
with width $d_v = 32$ and ReLU activation $\sigma$, where $P$, $W_l$ and $Q$ are affine. We denote the resulting map, in physical units, by $S^{
\mathrm{NN}}$.

\paragraph{Loss function and the optimizer}
Instead of the relative $L^2$ loss \eqref{eq:FNOloss}, we use the mean squared error of the normalized outputs,
\begin{equation}
    \mathcal{L} = \frac{1}{3N} \sum_{i =1}^{N}\| \tilde{y}^{\mathrm{NN}}-\tilde{y_i}\|_2^2, \qquad N = 4,200
\end{equation}
and minimize it with the Adam optimizer \cite{kingma2015adam} with learning rate $10^{-3}$ for 6,000 iterations (epochs), each a gradient step on all 4,200 training samples (full-batch), keeping the parameters with the lowest validation loss.

\subsection{Hybrid simulator}
The spring-slider version of Algorithm H is as follows.

\begin{mdframed}
{\bf Algorithm H (spring-slider)}

\begin{enumerate}
  \item \textbf{Interseismic (QD) phase:} integrate the system ODE until the event-start criterion is satisfied (event detected at $t=t_s^n$, $x(t^n_s) \in \Sigma_s$).
  \item \textbf{Seismic (surrogate) phase:} apply the learned jump operator to obtain the
  post-event state, $x(t_e^n)=S^\text{NN}\bigl(x(t_s^n)\bigr)$.
  \item Repeat Steps 1--2 until the desired number of events is reached.
\end{enumerate}
\end{mdframed}

\subsection{Results}

\begin{figure}\label{fig:spring-slider-sequence}
    \centering
    \includegraphics[width=0.75\linewidth]{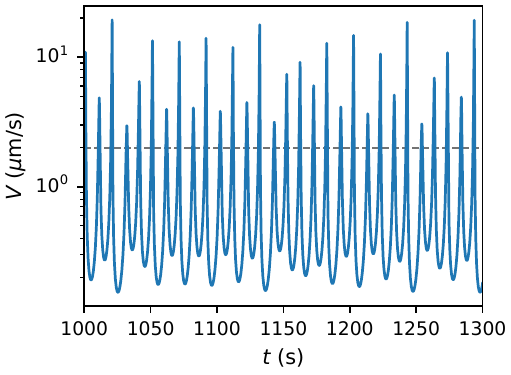}
    \caption{Evolution of slip rate in the spring-slider system simulation. The dashed line is the event threshold $ V=2V_0$.}
\end{figure}

Figure~\ref{fig:spring-slider-sequence} shows a typical slip-rate sequence of the spring-slider system in \eqref{ode}. The sequence is aperiodic, and the slip rate varies between 0.15 and 19.4 $\mu$m/s. We confirm the positive Lyapunov exponent, so the sequence is chaotic, in agreement with \cite{ShkollerMinster1997, Becker2000}.

\begin{figure}\label{fig:spring-slider-regression}
    \centering
    \includegraphics[width=1\linewidth]{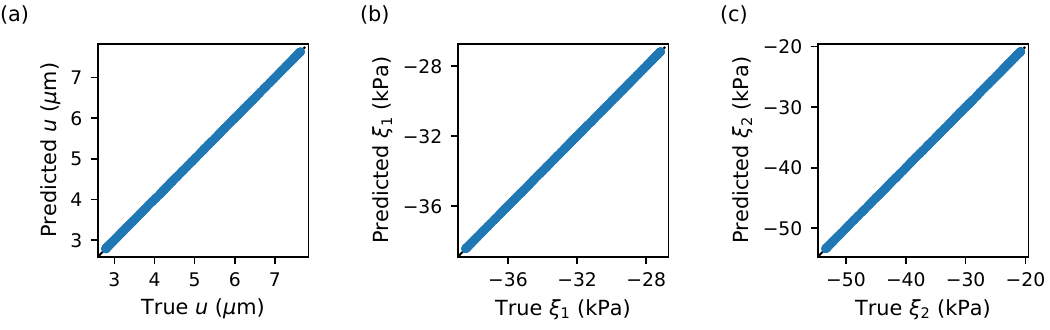}
    \caption{Predicted versus true output state $x(t^n_e)$ for the 1,200 test samples: (a) $u$, (b) $\Theta_1$, (c) $\Theta_2$.}
    \label{fig:placeholder}
\end{figure}

Figure~\ref{fig:spring-slider-regression} compares the predicted and true output states for the 1,200 test samples. The MSEs of the three normalized outputs are $4.7\times10^{-6}$ for $u$, $3.8 \times 10^{-6}$ for $\Theta_1$ and $4.7\times10^{-6}$ for $\Theta_2$.

\begin{figure}[h]\label{fig:spring-slider-phase}
    \centering
    \includegraphics[width=1\linewidth]{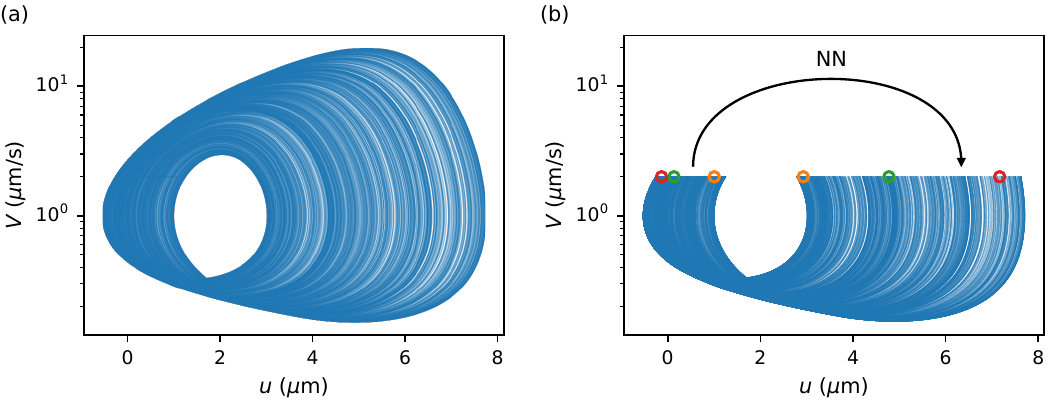}
    \caption{Comparison of (a) the ground-truth trajectory and (b) the hybrid trajectory in the $u$–$V$ phase plane. In (b), an NN mapping replaces the Poincaré map. The three colors indicate example input–output pairs, with matching colors identifying corresponding inputs and outputs.}
    \label{fig:placeholder}
\end{figure}

\begin{figure}[h]\label{fig:spring-slider-statistics}
    \centering
    \includegraphics[width=1\linewidth]{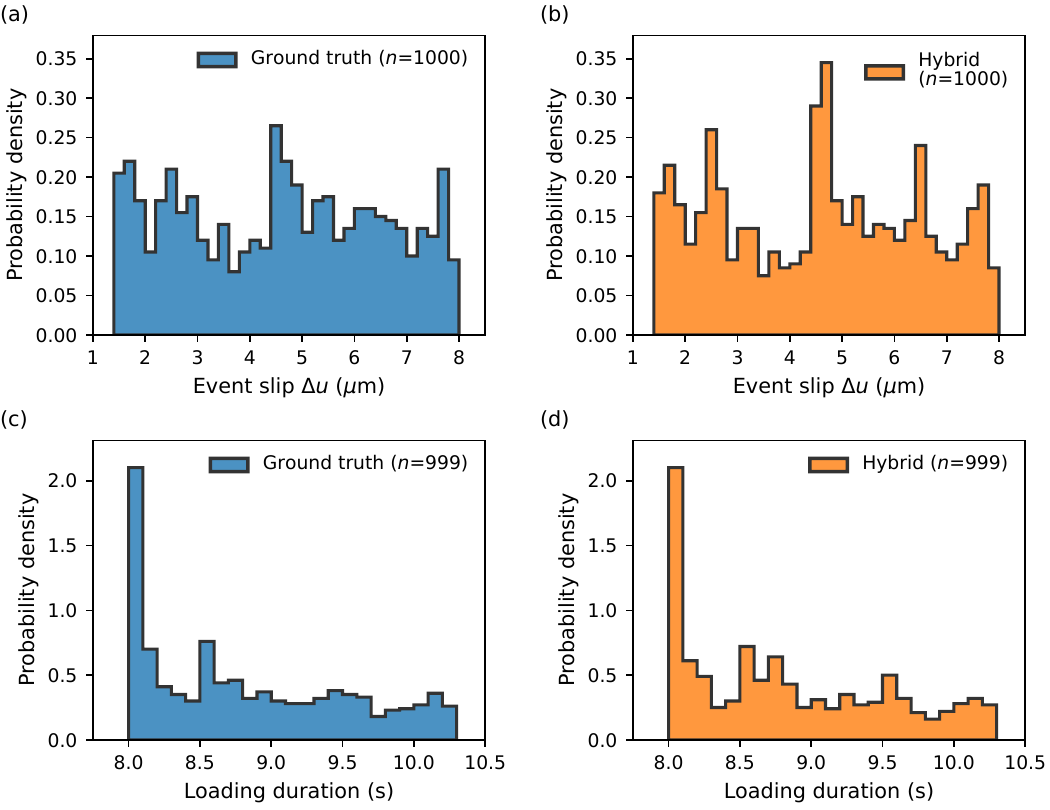}
    \caption{Distributions of event slip (a,b) $\Delta u_n$ and (c,d) loading duration $T^{\mathrm{load}}_n$ for the ground truth (left) and the hybrid simulation (right).}
    \label{fig:placeholder}
\end{figure}

With the trained NN map, we run the hybrid simulation for 1,000 events after the transient. The true trajectory and the hybrid simulation trajectories can be found in Figure~\ref{fig:spring-slider-phase}. Figure~\ref{fig:spring-slider-statistics} shows the distribution of event slip and loading duration. The $W_2$ distance and KL divergence between the distributions are summarized in Table \ref{tab:spring-slider-probdistances}. The hybrid simulation reproduces both statistics.

\begin{table}[h]\label{tab:spring-slider-probdistances}
    \centering
    \caption{Wasserstein distance $W_2$ and KL divergence between
    the hybrid and ground-truth event catalogs
    (1,000 events and 999 loading durations).}
    \label{tab:event-catalog-distances}
    \begin{tabular}{lcc}
        \toprule
        & $W_2$ & KL \\
        \midrule
        Event slip $\Delta u$
            & $0.0574\,\mu\mathrm{m}$ & 0.0620 \\
        Loading duration $T^{\mathrm{load}}$
            & $0.0220\,\mathrm{s}$ & 0.0685 \\
        \bottomrule
    \end{tabular}
\end{table}

\end{document}